%% file: ms.tex
\documentclass[10pt,preprint]{aastex}                                                                                                    

\shorttitle{Surface Detonations}
\shortauthors{Meakin et al.}

\def\nuc#1#2{\relax\ifmmode{}^{#1}{\protect\text{#2}}\else${}^{#1}$#2\fi}
\def\solrat#1#2{$[$#1/#2$]$}
\def\msol#1{\relax$#1\,M_\odot\/$ }
\def\mcol{\multicolumn}

\def\msun{M$_{\odot}$}

\def\tnm{\tablenotemark}
\def\tnt{\tablenotetext}
\def\iso#1#2{$^{#2}${#1}}
\def\betp{$\beta^+$}
\def\betm{$\beta^-$}

\begin{document}
\title{STUDY OF THE DETONATION PHASE IN THE GRAVITATIONALLY CONFINED DETONATION MODEL
OF TYPE Ia SUPERNOVAE}
\author{Casey A. Meakin\altaffilmark{1,2,3,5}, Ivo Seitenzahl\altaffilmark{3,4}, 
  Dean Townsley\altaffilmark{1,2,3}, George C. Jordan IV\altaffilmark{1,2},
  James Truran\altaffilmark{1,2,3}, Don Lamb\altaffilmark{1,2,4}}
\altaffiltext{1}{Center for Astrophysical Thermonuclear Flashes, University of Chicago, Chicago, IL}    
\altaffiltext{2}{Department of Astronomy and Astrophysics, University of Chicago, Chicago, IL}          
\altaffiltext{3}{Joint Institute for Nuclear Astrophysics, University of Chicago, Chicago, IL}          
\altaffiltext{4}{Enrico Fermi Institute, University of Chicago, Chicago, IL}                            
\altaffiltext{5}{Steward Observatory, University of Arizona, Tucson, AZ}                                
\email{casey.meakin@gmail.com}

\begin{abstract}
  We study the gravitationally confined detonation (GCD) model of Type
  Ia supernovae through the detonation phase and into homologous
  expansion.  In the GCD model, a detonation is triggered by the surface
  flow due to single point, off-center flame ignition in carbon-oxygen
  white dwarfs.
  The simulations are unique in terms of the degree to which non-idealized physics is
  used to treat the reactive flow, including weak reaction rates and a time dependent
  treatment of material in nuclear statistical equilibrium (NSE).
  Careful attention is paid to accurately calculating the final composition of material
  which is burned to NSE and frozen out in the rapid expansion following the passage of a 
  detonation wave over the high density core of the white dwarf; and an efficient method
  for nucleosynthesis post-processing is developed which obviates the need for costly network 
  calculations along tracer particle thermodynamic trajectories.  Observational diagnostics are 
  presented for the explosion models, including abundance stratifications and integrated yields.
  We find that for all of the ignition conditions studied here, a self regulating process comprised 
  of neutronization and stellar expansion results in final \iso{Ni}{56} masses of $\sim$1.1\msun.
  But, more energetic models result in larger total NSE and stable Fe peak yields.
  The total yield of intermediate mass elements is $\sim0.1$\msun and the explosion energies are 
  all around 1.5$\times$10$^{51}$ ergs. The explosion models are briefly compared 
  to the inferred properties of recent Type Ia supernova observations.
  The potential for surface detonation models to produce lower 
  luminosity (lower \iso{Ni}{56} mass) supernovae is discussed. 
\end{abstract}


\section{INTRODUCTION}

\par The currently favored model for Type Ia supernovae (SNe Ia) is the 
thermonuclear incineration of a white dwarf (WD) which has accreted mass
to near the Chandrasekhar limit from a binary companion 
\citep[e.g.,][]{branch1995,hillebrandt2000}. The enormous luminosity
and homogeneity in the properties of the light curves of SNe Ia make
them exceptionally good standard candles and as such have shown that the
expansion rate of the universe is accelerating and provided intriguing evidence
for a cosmological constant \citep{riess1998}.

\par Despite the success in using SNe Ia as cosmological probes and identifying
a plausible astrophysical progenitor site for the explosions, a detailed
understanding of the explosion mechanisms itself remains elusive. 
Several uncertainties stand in the way of a definitive solution to the
SNe Ia problem. On the one hand, the conditions under which the 
thermonuclear runaway commences remains poorly understood so that the
initial number and distribution of flamelets that seed the runaway
is still a free parameter. On the other hand, although significant progress has been
made in simulating flame fronts in multi-dimensional stellar models
\citep{gamezo2005,schmidt2006b,roepke2007a,townsley2007,jordan2007}, the challenge
associated with modeling an unresolved turbulent deflagration 
\citep[e.g.][]{schmidt2006a} with limited computational resources injects
an additional degree of uncertainty into the outcome of a model for any 
given choice of initial conditions.

\par In this paper we describe progress on our ongoing effort to improve
the simulation of SNe Ia in multi-dimensions, including methods
to perform detailed nucleosynthesis post-processing in a computationally efficient manner. 
We extend the study of the GCD
model for a single ignition point slightly offset a range of distances from the center of the
star, as described in \citet{townsley2007}, through 
the detonation phase and into homologous expansion.
In \S\ref{sec:numerics} we describe the treatment of the reactive-hydrodynamics 
problem used in our simulation code. In \S\ref{sec:def} we review the relevant properties of 
the deflagration phase for single point flame ignition.  In \S\ref{sec:det} we examine 
in some detail the initiation of the detonation, the properties of the detonation 
wave which disrupts the star, and the resultant remnant morphology. 
In \S\ref{sec:yields} we discuss in detail the nucleosynthetic yields for the explosions
studied and decribe the methodology used to efficiently calculate iron peak yields
from the simulation data. We conclude with a summary of the salient features of the 
explosion models in light of observed Type Ia supernvoae.

\section{NUMERICAL METHODS: HYDRODYNAMICS AND NUCLEAR BURNING}
\label{sec:numerics}
\par In this section we review the computational tools used to model the hydrodynamic 
and nuclear evolution of the stellar plasma, including the treatment of subsonic 
(deflagration) and supersonic (detonation) burning fronts.  The basic code framework is 
FLASH \citep{fryxell2000}, a modular, block-structured adaptive mesh refinement (AMR), 
Eulerian, reactive-hydrodynamics code. We use a directionally split PPM solver \citep{colella1984} 
generalized to treat non-ideal gasses \citep{colella1985} to handle the hydrodynamic 
evolution.

\par The energetics scheme used to treat flames and detonation waves
in our simulations uses 3 progress variables to track carbon burning, 
NSQE relaxation, and NSE relaxation.  The rates connecting these burning 
stages are calibrated using a large (200 nuclide) nuclear reaction network 
for the conditions relevant to the Type Ia problem. Additionally, energy losses 
(through neutrino emission) and changes in the electron mole fraction
$Y_e$ due to weak interactions taking place in material which has burned to NSE are
incorporated.  Details can be found in \citet{calder2007,townsley2007,seitenzahl2008a}.
%\par Nuclear burning is coupled to the hydrodynamic evolution using an operator splitting 
%formulation whereby hydrodynamic and nuclear burning modules are called succesively over 
%the course of a single timestep \citep{fryxell2000}.

\par Both detonation waves and flames are impossible to resolve in full star simulations
because they are characterized by length scales that are more than ten orders of magnitude 
smaller than the radius of the white dwarf to be modeled, $R_{\rm wd}\sim10^8$ cm.  
Therefore, these reaction fronts must be treated in a special manner. Subsonic burning fronts 
(deflagrations) are advanced using an advection-diffusion-reaction (ADR) equation.
In short, a thickened flame front  ($\sim$4 grid zones wide) is advanced at a 
speed $v_f = \max(v_l, v_t)$, where $v_l$ is the laminar flame speed calculated 
by \citet{timmes1992} and  $v_t$ is a Rayleigh-Taylor driven 
turbulent flame speed. Details concerning the implementation, calibration and noise 
properties of the flame treatment can be found in \citet{townsley2007} and 
Asida et al. (2008, in preparation) and references therein.

\par Detonations are handled naturally by the reactive hydrodynamics solver
in FLASH without the need for a front tracker. This approach is possible because 
unresolved Chapman-Jouguet (CJ) detonations retain the correct jump conditions and propagation speeds. 
Numerical stability is maintained by preventing nuclear burning within the
shock.  This is necessary because shocks are artificially spread out over a few zones 
by the PPM hydrodynamics solver, which can lead to unphysical burning within
shocks that can destabilize the burning front \citep{fryxell1989}.
The energetics in the detonation differ from that in the deflagration front only
in how carbon burning proceeds, as represented by the first progress variable $\phi_1$
and an explicit carbon burning rate is used \citep{caughlan1988}. The additional burning stages
(NSQE and NSE relaxation) are tracked by $\phi_2$ and $\phi_3$ and are evolved in the same 
manner as in the post flame ash \citep{calder2007,townsley2007}.
At densities above $\sim 10^7$ g cm$^{-3}$ 
detonations propagating through a mixture that is equal parts \nuc{12}{C} and \nuc{16}{O} have 
Mach numbers that are larger than the CJ value, but by only a few percent 
\citep{gamezo1999,sharpe2001}.
Cellular structure smaller than the grid scale will be suppressed in our simulations 
but is free to form on resolved scales. The impact of cellular structure on the global evolution
of the model is still uncertain.
However, since cellular structure alters the detonation wave 
speed by only a few percent for the conditions being modeled \citep{timmes2000b} the effect is likely to be small. 
Additional details related to the treatment of detonation waves are discussed in \S\ref{sec:det}.

%In order to treat detonations in the explicit manner described, a carbon 
%burning reaction rate is needed \citep{caughlan1988} to advance the first
%progress variable in place of the flame model. The additional burning stages 
%($\phi_2$, $\phi_3$) are advanced using the relaxation timescales in the same 
%manner as in the flame.

%\par This approximate treatment of the nuclear burning has the following shortcoming: 
%For densities above $10^7$ g/cm$^3$ detonations propagating through a mixture that 
%is equal parts \nuc{12}{C} and \nuc{16}{O} are in the ``pathological'' regime 
%\citep{gamezo1999,sharpe2001}. [Basic features: detonation speeds vary less 
%than X\% over the region of interest.]  
%Detonation waves are unstable to perturbations transverse to the direction 
%of propagation which give rise to a so-called ``cellular structures'' on length 
%scales comparable to the induction length \citep{gamezo1999,timmes2000b}.  
%Suppresed on smaller scales by resolution -- what might be the subgrid-scale 
%effect of these instabilities? clumpiness, propagation speed differences. 
%(Summarize. Conclude by reciting the magnitude of the errors expected. $\sim$ a few \%)

\par Self gravity is calculated using a multi-pole solver with a maximum spherical harmonic
index l$_{max}$=10. The Helmholtz equation of state of \citet{timmes2000a} is used to 
describe the thermodynamic properties of the stellar plasma including contributions from 
blackbody radiation, ions, and electrons of an arbitrary degree of degeneracy.

\par Passive tracer particles are included in our simulations which record
the time history of the flow properties along Lagrangian trajectories. These records
can be used to calculate detailed nucleosynthetic yields as well as to provide
additional diagnostic for complex flows.  We use 10$^5$ tracer particles
for 2D models and 10$^6$ for 3D models.  The particles used in this study are
initialized at the beginning of each simulation with a mass weighted distribution.
In \S\ref{subsec:freezeout-method} we present a novel method to calculate post-explosion
yields which does not require the prohibitively expensive post-processing of a large
number of tracer particle with a nuclear reaction network, but rather uses information 
readily extracted from the tracers to calibrate an efficient table look-up scheme.

\section{FLAME IGNITION AND DEFLAGRATION}
\label{sec:def}
\par In this paper we extend the study of the GCD model for the single point flame ignition models of
\citet{townsley2007} through the detonation phase and into homologous
expansion. The general simulation
setup is the same, and we review it here briefly, along with a description of
the basic progression of the evolution preceding detonation.
After carbon burning ignites at the center of the white dwarf, a
convective core is formed which expands as it heats.  Our simulations begin
when the nuclear burning timescale becomes shorter than the eddy turnover
time, so that the first flamelet is ignited near (within a few hundred km of)
the center of the white dwarf.  As discussed by \citet{townsley2007}, there is
still significant uncertainty
in the form which the nuclear flame will take at birth (also see
e.g.~\citealt{woosley2004}), relating to the number and location of what are
generally assumed to be relatively small ($<1$ km) ignition regions.  For
reasons related to simplicity of setup and limitations of the imposed
cylindrical symmetry, we restrict our study to off-center, single point
ignitions in a quiescent background star.
The initial WD used in these simulations has a uniform temperature $4\times
10^7$ K, a mass of \msol{1.365}\footnote[1]{This was erroneously given as
\msol{1.38} in \citet{townsley2007}, none of these parameters have
changed from that work to this.}, a central density of $2.2\times 10^9$ g
cm$^{-3}$, and is composed of equal mass fractions of \nuc{12}{C} and
\nuc{16}{O}.  This progenitor is much colder than reality, but it is
expected, and we have confirmed by comparison, that this does not have a
significant effect on the structure of the white dwarf or the dynamics 
of the explosion.  At the beginning of the
simulation a spherical region of radius $r_{\rm bub}$=16 km placed on
the polar axis at a range of distances between 
$r_{\rm off}=$20 and 100 km from the center of the star is converted to NSE 
ash in pressure equilibrium with the remainder of the star.  A summary of the 
initial flame bubble parameters studied in this paper is given in 
Table \ref{tab:models}.

\par The basic stages of single bubble flame evolution can be described in
terms of two key length scales, the grid resolution, $\Delta$, which sets the
limit to which we can resolve flame structure, and the critical wavelength,
sometimes called the fire polishing length, $\lambda_c\equiv 6\pi s^2/Ag$ 
\citep{khoklov1995}, where $s$ is the front propagation speed (flame speed)
and $A$ is the Atwood number 
$A=(\rho_{\rm fuel}-\rho_{\rm ash})/(\rho_{\rm fuel}+\rho_{\rm ash})$.  
Perturbations in a
Rayleigh-Taylor (R-T) unstable flame front smaller than $\lambda_c$ are
``polished out'' by the propagation of the flame,  while those of larger
scale are enhanced by R-T growth, wrinkling the flame bubble.
\cite{townsley2007} distinguished three phases of flame evolution that occur
successively as the buoyant flame bubble grows in size and rises toward the
stellar surface: laminar bubble growth when $r_{\rm bub} \lesssim\lambda_c$,
resolved R-T unstable growth, and finally R-T unstable growth on the sub-grid
scale when $\lambda_c < \Delta$.  One immediate consequence of this progression 
and the increase of $g$ with radius, is that the resolution limits the largest
ignition offset position $r_{\rm off}$ which can be used and still start in 
the laminar growth phase. This limit is roughly 100 km for $\Delta=4$ km, the 
resolution of all the simulations here. At this resolution the critical wavelength
is less than the grid scale, $\lambda_c < \Delta$, outside roughly 400 km from the 
center of the star.

\begin{figure}
  \includegraphics[scale=0.75]{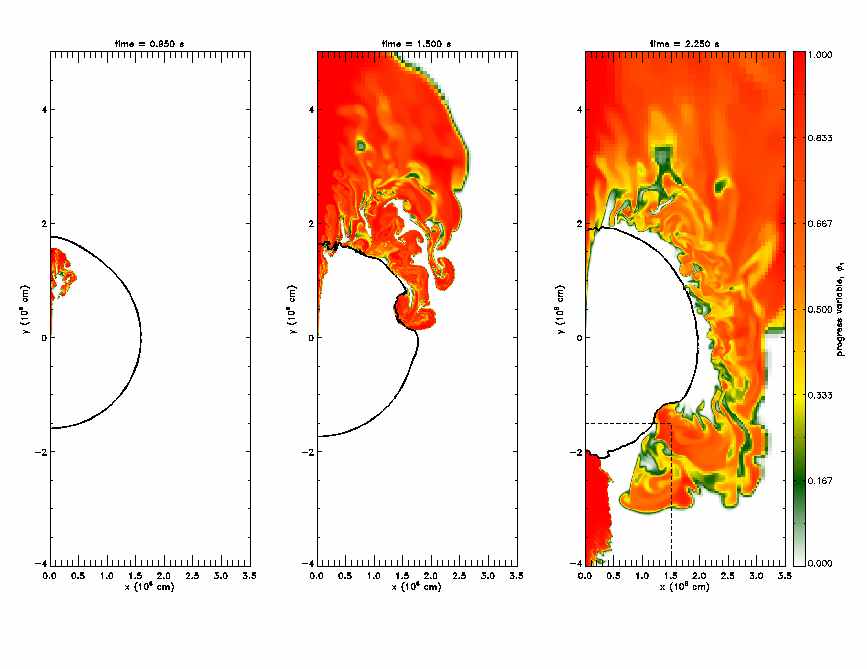}
  \caption{\label{fig:breakout}
  This time sequence of ash abundance (represented by the $\phi_1$
  progress variable) shows how a bubble ignited near the stellar core rises
  buoyantly, erupts from the star's surface and drives a flow which is largely
  confined to the surface of the star by gravity.  The black contour line indicates 
  a density of 10$^7$ g cm$^{-3}$.  Eventually the surface flow converges at the 
  opposite pole from the breakout location, compressing material in that region
  until it begins to burn carbon. The dashed box in the right panel indicates
  the region detailed in Figure \ref{fig:jet} below, where the converging
  flow produces a jet that initiates a detonation wave.}
\end{figure}

For even modest offsets, the hot ash is buoyant enough that it erupts from
the surface of the star before more than a few percent of the star is burned.
This creates a vigorous flow over the surface of the WD, which is still
relatively compact due to the small amount of burning.  The progress of this
eruption and flow is shown in Figure \ref{fig:breakout} for the case with
$r_{\rm off} = 40$ km.
As mentioned previously, there is indication from comparisons of recent work
\citep{roepke2007a,townsley2007,jordan2007} that the eruption pattern arising
from a given offset is dependent on the choice of burning model, and therefore is
currently uncertain.  This has important consequences at the collision
region because it sets the velocity and density structure of the
incoming flows as well as the surface gravity (via the degree of stellar
expansion) under which the collision occurs.

We study the cases from \citet{townsley2007}, which demonstrated collisions
that created detonation conditions, along with several supplementary cases near the
minimum offset distance that led to detonation conditions.  This gives the range $r_{\rm
off}=$20 to 100 km.  The expansion which occurs during the
deflagration and surface flow stages is very nearly homologous and has only a
small degree of asymmetry, such that most of the asymmetry is created later
during the detonation phase (see below).
Figure \ref{fig:homologous-deflagration} compares the scaled density profile
for the initial model and the 25 km and 100 km offset cases at the time the
detonation initiates.  Density is scaled by the central value and radius is
scaled by the distance from the center at which the density drops to $1/e$
times the central value.  Profiles along the equator and along the symmetry
axis are both shown, demonstrating that the star remains very symmetric out
to approximately 2 density scale heights away from the center.  This region
contains approximately 90\% of the stellar mass, and even the asymmetry
beyond this is fairly modest, but is likely to lead to some asymmetry in the
highest-velocity spectral features.

Based on the variation in expansion found in the resolution study performed
by \citet{townsley2007}, and other cases generally, it appears that the
conditions at the collision region and the density structure at detonation
are not a single parameter family in the mass burned, or equivalently nuclear
energy release, in the deflagration, $E_{\rm n,def}$, prior to the collision.
It does seem that $E_{\rm n, def}$ is the primary parameter, but other
contributing factors include the time dependence of the energy release,
morphology of the burning region in the flame plume, and the possibility of
secondary or tertiary ignition sites.  The sudden input of energy in the
deflagration puts the star in an oscillation, and the timing of the
detonation initiation with respect to this oscillation is important for
setting the density structure at detonation, and thereby the burning
products (see \S5).  The timing and magnitude of the
nuclear energy release will change the magnitude and phase of this full-star
oscillation, but further investigation, including studies of 3-dimensional
deflagration morphologies, is needed to characterize these relationships.

As an aside we note the impact of neutronization due to electron captures
during the deflagration.  Neutronization influences the dynamics of a rising
flame bubble by changing the average binding energy of the final NSE state
obtained as well as the electron pressure available per gram of stellar plasma. 
We assessed the impact of neutronization by recalculating several models through 
bubble rise with weak rates suppressed, by enforcing $\dot Y_e = 0$. Flame bubbles 
ignited closer to the stellar center, and hence at higher densities,
are more strongly affected. Models ignited at $r_{\rm off}$=40 km and 30 km 
{\em burned $\sim$10\% and $\sim$40\% more mass, respectively, with weak rates 
suppressed} while the model ignited at $r_{\rm off}=$80 km 
was negligibly affected by the weak reactions during bubble rise and breakout.
The effect that the weak reactions have on the burned mass depends on the
developement of turbulence which is not well represented in the 2D simulations
presented here.  Therefore, while we have demonstrated that weak reactions play
a non-negligible role in the present suite of models the impact that they have
on more realistic 3D flows remains an open question.

%(Based on a comparison with Dean's earlier runs to my more recent ones
% w/ and w/out weak rates it appears that the sensitivity of the calculations
% to the development of turbulence is very strong and a simple explanation of the
% differences remains lacking.)
%
% It appears 
% that the change in the binding energy of the final state dominates such that flames with 
% weak reactions suppressed release less energy in burning to NSE thereby producing less 
% buoyant ash which rises more slowly, burns to a larger lateral extent, and produces 
% more total burned mass prior to breakout.
%
% \par (Dean to Casey:  This is not consistent with the studies I did on this, in
% which the bubble rise was directly observed to be faster with neutronization
% turned off.  I didn't check the burned mass dependence.  It may be that the
% faster rise generates more turbulence...)

\section{DETONATION}
\label{sec:det}

%\par Even slightly off-center flame ignition in the core of white dwarf appears
%to strongly diminish the chances that a deflagration by itself can unbind a
%white dwarf and lead to an explosion comparable to those observed as Ia SNe. 

%\par While the conditions of flame ignition, including the spatial distributions and formation
%rate, remains unknown an increasing  number of studies have begun to explore the mapping 
%between proposed distributions and final outcomes, incorporating 
%ever more realistic physics and relying increasingly on 3D full star simulations
%\citep{gamezo2005,jordan2007,roepke2007a}.
%Although the correct treatment of a turbulent flame remains an active area of research,
%recent simulations which use state of the art turbulent flame models 
%cannot reproduce the observed features of normal luminosity Ia SNe when burning takes 
%place solely in a deflagration \citep{roepke2007b}. These results have renewed interest 
%in scenarios which undergo a transition to detonation after a phase of subsonic burning.

\par Single point off-center ignition results in a buoyant plume of hot ash which is brought 
to the surface of the star before more than a few percent of the stellar core is consumed 
by the flame (see M$_{\rm def}$ in Table~\ref{tab:models}). As the hot ash rises to the surface, 
the nuclear energy that is released excites a stellar pulsation which initially expands the star.  
Against this background pulsation, 
the hot ash from the burning is expelled from the stellar interior. A large fraction of this ash 
is confined to the star's surface by gravity.  This ash sweeps over the surface of the
star together with a flow of unburned stellar material which is pushed ahead of it.
In all but the most expanded model in our parameter study (i.e., those with ignition points
r$_{\rm off} \ge$ 25 km), the resulting surface flows converge at a point opposite to breakout 
which we refer to as the {\em collision region}  (Fig.~\ref{fig:breakout}). 
These converging surface flows result in a bi-directional, collimated jet-like 
flow which both expels material away from the star's surface and drives a flow of high 
temperature material into the stellar core.
The inward directed component of the collimated flow 
reaches high enough densities and temperatures that a ``surface detonation'' inititiates which 
sweeps over the core and completely disrupts the white dwarf, giving rise 
to a luminous supernova explosion. 

\par In the following subsections we describe the
characteristics of the bi-directional jet which forms in the colliding surface 
flow and initiates the detonation (\S\ref{subsec:jet}), 
we discuss the characteristics of the ensuing detonation phase of burning (\S\ref{subsec:det}), 
and we describe the final state of free expansion which results (\S\ref{subsec:remnant}).
A detailed analysis of the nucleosynthetic yields is presented in the next section, \S5.

%Only the model with an initial flame bubble 
%ignited 20 km off center fails to detonate within the time simulated. In this model the large 
%amplitude of the stellar pulsation produced by the energy release in the deflagration 
%stalls the surface flow and lowers the densities in the collision region to values 
%well below the canonical value of $\rho\sim$10$^7$ g cm$^{-3}$ necessary to trigger a 
%detonation \citep{niemeyer1997,seitenzahl2008b}.

\subsection{Jet Formation and Detonation Initiation}
\label{subsec:jet}

%% FORMATION AND INIITAL DEVELOPMENT OF THE JET

\par {\em Jet Formation and Characteristics.---} As the surface flow produced by the deflagration 
converges, material accumulates in a small region on the hemisphere opposite to the 
breakout location.  The material which initially piles up, consists of unburned
carbon and oxygen rich surface material which is pushed ahead of the ash as it flows around
the stellar surface.  As material accumulates in this region it is heated by compression until it reaches
temperatures sufficient to initiate carbon burning, which further heats the compressed material
and raises its pressure.  Shortly after the initial collision, a conical shock forms which separates the 
compressed material along the axis from the inflowing surface flow. The surface flow material
burns as it passes through this shock and ``accretes'' into the collision region.  
The resultant pressure in the collision region roughly
balances the ram pressure of the accreting surface flow, $p_{\rm coll} \sim [\rho v^2]_{\rm surf}$ (see Fig.~\ref{fig:jet-slice}).
The pressure achieved in the compressed region soon exceeds the (nearly hydrostatic) background pressure 
sufficiently that it redirects the accreting material and drives a bidirectional jet-like 
flow which has components aligned along the polar axis.  A closeup of the collision region 
thus formed is shown in Figure~\ref{fig:jet} (which corresponds to the region outlined
by the dashed box in Figure~\ref{fig:breakout}). The velocity vectors reveal the 
bidirectional nature of the flow. The ash from the deflagration is just approaching 
the collision region at the time shown, well after the collimated jet has formed.
The width of the jet increases with time as material continues to accrete into
the region, but retains structure on scales $<$50 km, which are well resolved in our
simulations which have a grid resolution of 4 km.

%% JET PROFILES

\par In Figure~\ref{fig:jet-slice} the flow properties along the jet axis are shown just prior to 
the onset of detonation for two 2D models and one 3D model.
The 2D models shown span the conditions studied in this paper, including the model with the most
expanded core (left-panel) and the least expanded core (middle-panel) which
detonate in our study.
All of the collimated flows share the same overall structure with
the more expanded stars having shallower density gradients in the collision region.
The velocity profiles are roughly linear, decreasing from a maximum inwardly
directed velocity of $\sim$ (1 to 2)$\times 10^9$ cm/s to a comparable
velocity directed away from the stellar surface.
While the inward flow is attended by a great deal of small scale internal 
substructure and turbulence (Fig.~\ref{fig:jet}), three distinct ``fronts'' are readily identifiable 
along the axis of the jet: a leading subsonic compression wave, followed by a fuel-ash boundary layer,
and finally an internal shock.  The fuel-ash boundary layer is marked in Figure~\ref{fig:jet-slice}
by the dashed vertical line. The material ahead of this line has not yet been compressed
to high enough densities that carbon burning can proceed.
The compression wave(s) which eminates from the head of the jet as it moves
into the star can be seen as perturbations preceeding the fuel-ash boundary in all of the variables 
plotted in Figure~\ref{fig:jet-slice} and can also be seen as the pressure waves extending into
the star ahead of the jet in Figure~\ref{fig:jet} (left-panel).  The compression 
wave moves into the star at the sound speed, which is 
$c_s\sim 3.5\times 10^8$ cm/s at this location. The head of the jet, as marked by 
the location of the fuel-ash boundary, moves inward at a fraction of the sound speed so that 
the size of the compressed region grows with time. 
Trailing behind the fuel-ash boundary is a shock wave which
separates the low density, high velocity flow produced in the collision region from the compression
wave which moves ahead of it. It is the ram pressure of this high velocity flow which drives the 
compression wave into the star.  The ram pressure of this high velocity flow is balanced by the 
gas pressure of the compressed, overlying material, as shown by the red line in the bottom panels of 
Figure~\ref{fig:jet-slice}.  In all of the models studied, the inward directed jet continues to
compress material, heating it to carbon burning conditions until a detonation arises
and distrupt the star.

\par An important question in the context of the present study concerns the extent to which the jet-like 
flows which develop depend upon the 2D geometry used.
Simulations of off-center ignition using 3D grids have been made \citep[e.g.][]{roepke2007a,jordan2007} 
with the general conclusions that focusing of surface flow also occurs in 3D and is not strongly 
diminished compared to 2D. As a point of direct comparison, we have simulated a 3D model from flame
ignition through detonation using the same methods as used in the 2D models presented here.  
This 3D model used a finest resolution of $\Delta$=8 km and was ignited
by a 16 km flame bubble displaced 80 km from the stellar center.  The development of the collision 
region and the subsequent detonation in the 3D model is remarkably similar to that found in the 2D models.  
For comparison, the profile of the jet formed in the 3D model just prior to detonation is included in 
the right panel of Figure~\ref{fig:jet-slice}.

%% SHAPED CHARGE JETS

\par Jet formation within a converging flow and jet penetration are well studied phenomena.  
For instance, engineers have designed shaped charge explosives which create jets by explosively 
collapsing a convex ``liner'' material, most often cone-shaped, onto itself \citep[e.g.][]{birkhoff1948}.\footnote[2]{Mining engineers have employed similar methods as early as 1792.}
The jets thus formed have notoriously strong penetrative power and can
slice and penetrate sheets of steel which are several times thicker than
the shaped charge diameter and are applied in both military and industrial 
capacities including metal perforation, armor penetration, and oil well drilling.
While there are many differences between shaped charge jet formation and 
the collision region flows present in our calculations, the phenomena bear interesting 
similarities which may provide insight into the depth to which a converging surface flow may
penetrate the underlying carbon-oxygen rich layers of a white dwarf.  
The jet models which have been 
made to interpret experimental results estimate penetration depth by balancing 
the ram pressure of the jet material with that of the target material in a frame of reference 
that is moving with the jet-target interface. The penetration depth under the simplifying 
circumstances of constant density jet and target materials depend on only the density ratio and
the jet length.  This picture is greatly complicated in the stellar surface flow
case where compressibility plays a central role and the pressure balance at the jet-star interface 
is between the dynamical pressure of the jet and the gas pressure of the core. 
While it is
beyond the scope of the present paper to fully analyze the problem of compressible jet formation and 
penetration, we conclude by noting that the strong penetrative power observed in shaped charge jets
provides support for the deep penetration
seen in all of the simulations in our study which develop collision regions.  In all of the cases
which we study, the jets which have formed penetrate into denser layers of the white dwarf until a 
detonation occurs.

%% THE INITIATION OF THE DETONATION

\par {\em Detonation Initiation.---} Once the density of the material undergoing carbon burning
in the jet exceeds $\rho_{\rm det}\sim 10^7$ g/cm$^3$ a detonation initiates 
which then propagates away from the head of the jet at the Chapman-Jouguet speed,
$D_{CJ} \sim 1.2\times 10^9$ cm/s with Mach number $M = D_{CJ}/c_s \sim 3.4$.
The time sequence shown in Figure~\ref{fig:jet} captures the moment when the 
detonation initiates at the head of the jet and begins to spread outward. Because
of the weak dependence of the detonation wave speed on the upstream density, the detonation 
front radiates from its point of initiation nearly spherically. 

\par The initiation of the detonation, which takes place at the fuel-ash boundary, when 
$\rho\sim 10^7$ g/cm$^{3}$ and T$\sim 3\times10^9$ K., resembles 
a Zel'dovich gradient mechanism \citep{zeldovich1970,khokhlov1997}.  Detonation initiation through this process
involves a complicated interplay between burning and hydrodynamic flow
that requires a coherent build up of acoustic energy by the nuclear energy release.  
An often cited criteria for the initiation 
of a detonation in the context of degenerate carbon-oxygen material is that a ``critical'' 
mass of material needs to be heated and compressed above a certain temperature and density 
threshold \citep{arnett1994,niemeyer1997,roepke2007a}.  While these studies indicate the general
conditions under which detonations might readily arise, thermodynamic
conditions and heated masses alone represent a gross oversimplification of the underlying initiation process which
depends sensitively on the gradients of thermodynamic variables within the heated region.  
Since gradients play a central role, the resolution and the geometry of the flows being simulated,
such as those presented here, are important considerations when investigating the potential 
for detonation. The suite of simulations studied in this paper use a finest zone size which is 4 km 
and limits the steepness of temperature gradients which can be represented in our models.
And although detonations do arise in our simulations, drawing conclusions from the results of
simulations alone concerning the success or failure of detonation will require investigations at 
significantly higher resolution than has been possible to date.

\par We have made some efforts to address the robustness of initiation with a suite of simulation 
models which employ a patch of mesh refinement over the collision region having zones as fine as 
125 m. One of the principal findings of this study, which is being prepared for publication
elsewhere (C. Meakin et al. in prep), is that the gradients at the head of the inward directed
jet component become steeper at higher resolution which at first appears to inhibit detonation. 
However, the higher resolution flows develop turbulent structures within the shear layers that 
form at the interface between the head of the jet and the background stellar material, 
such as through the Kelvin-Helmholtz instability, which thicken the fuel-ash boundary to an extent
that induction time gradients conducive to the spontaneous initiation of a detonation may develop after all.

\subsection{Propagation of the Detonation Wave over the Stellar Core}
\label{subsec:det}

\par Once the detonation wave forms it propogates outward from the spot 
of initiation nearly spherically, and consumes the unburned carbon and oxygen remaining 
in the core. The time sequence in Figure~\ref{fig:det-temp} shows the geometry of the detonation 
wave as it propagates over the stellar core. 
The detonation wave speed is a weak function of the upstream plasma density and varies by 
only $\pm$5\% for the conditions present in the uburned core, where $10^7<\rho<10^9$ g cm$^{-3}$ 
\citep[see Figure 2 of][]{gamezo1999}. The detonation wave traverses the expanded white dwarf 
in $t_{\rm cross}\sim 2 r_{\rm det}/{D_{CJ}}\sim 0.4$ s where the core size is roughly 
$r_{\rm det}\sim 2\times10^8$ cm and the detonation wave speed is $D_{CJ}\sim10^9$ cm s$^{-1}$.

\par As the detonation wave propagates it compresses upstream material prior
to burning.  Upstream material with a density greater than $\sim 10^7$ g cm$^{-3}$
is compressed and heated strongly enough by the shock that complete relaxation to nuclear 
statistical equilibrium (NSE) occurs before the rarefaction wave behind the detonation expands
the material and it freezes-out (see \S5). At lower upstream densities
relaxation to NSE is incomplete and the ash is composed of intermediate
mass elements (IMEs) such as Si, S, Ca, and Ar, i.e., the products of incomplete
silicon burning \citep[e.g.][]{woosley1973,arnett1996}.

\par Material which is compressed to densities exceeding $\sim10^8$ g cm$^{-3}$ in
the detonation wave develops a non-negligible neutron excess through electron capture 
reactions. The strong density dependence of the weak reaction rates limit this
neutronization to the central-most regions of the star as evident in 
Figure~\ref{fig:det-ye} which shows the spatial distribution of electron mole
fraction $Y_e$ as the detonation wave sweeps over the stellar core.  As discussed in \S5, the
final composition of the material burned to NSE, including the fraction which is \nuc{56}{Ni}, 
depends on the degree of  neutronization.

\par Detonation waves are subject to transverse instabilities which influence
the structure of the reaction zone and the reaction products and introduce
inhomogeneities in the downstream flow \citep[e.g.,][]{gamezo1999,timmes2000b,sharpe2001}.
Therefore, in order to faithfully capture in entirety the properties of the burning in a detonation 
wave the reaction length scale must be resolved.  An additional complication arises
in modeling detonations when the density scale height in the medium through which the detonation
propagates is comparable to or smaller than the reaction length.  Under these
conditions steady detonation wave theory cannot be applied and the resulting reactive-
hydrodynamic flow remains an active field of research \citep{sharpe2001}.  
In the context of a carbon-oxygen, near Chandrasekhar-mass white
dwarf (M$_{\rm Ch}$), such conditions arise when the upstream density is $\sim10^7$ g cm$^{-3}$.
Significant deviations from a Chapman-Jouguet detonation
may arise and influence the resulting intermediate mass element (IME) yield.  
Since IMEs, such as Si and Ca, are 
primary observational diagnostics of the explosion mechanism underlying SNe Ia 
\citep[e.g.][]{wang2003,wang2007}, these uncertainties have important implications
for modeling all delayed detonation scenarios.

\par In the models presented here, the stellar cores undergo only modest expansion
during the deflagration and detonation phases.  Between 90\% and 97\% of the unburned mass in the core 
has a density which exceeds $\sim10^7$ g cm$^{-3}$ at the time detonation initiates and 
all of this material undergoes complete relaxation to NSE, 
resulting in primarily \nuc{56}{Ni} and and a small fraction of stable Fe-peak elements 
(\S5 and Table~\ref{tab:models}).  Therefore, only a small amount of mass, which is confined to a thin 
shell in the outer part of the core, is burned to IMEs by the detonation wave.
Within this narrow shell the length scales associated with transverse
instabilities exceed the grid scale used ($\Delta$ = 4 km) \citep{gamezo1999} and our
numerical methods are sufficient to capture them.
However, material in this narrow region undergoes rapid expansion 
after the detonation wave passes and it quickly mixes with the turbulent layer of deflagration 
ash which lies immediately above it so that it is difficult to discern the presence of cellular 
structure if it did indeed arise.  Significantly higher fidelity simulations are required
in order to study the impact that transverse instabilities have under these conditions.
While these affects are negligible in the present suite of models, more expanded, lower 
density cores are likely to be much more strongly impacted by this uncertain physics.

\par Upon encountering the deflagration ash which enshrouds the star, 
the detonation wave transitions into a shock wave which accelerates
the hot ash.  After the detonation wave and the ensuing shock have propagated
off of the computational grid, what is left behind is a rapidly expanding
remnant consisting of a smoothly layered core of material burned to NSE with 
a thin shell of IMEs outside of that, surrounded by a turbulent
layer of ash from the deflagration composed of both NSE and
IME material.

\subsection{Transition to Free Expansion and Final Remnant Shape}
\label{subsec:remnant}

\par As the detonation wave traverses the stellar core it shifts the density
distribution so that the peak in density is initially moved in the positive 
$y$-direction. This can be seen in Figure~\ref{fig:det-propagation} which
presents a time series of velocity and density profiles spanning the time interval
over which the detonation wave traverses the stellar core.  The initial shift in the
density peak towards positive $y$ is due to the strong rarefaction which follows the detonation wave
and expands the material behind it on a very short timescale ($\tau_{\rm expand}\sim 0.4$ s).
Within $\sim$1 s following the passage of the detonation over the core, the density
peak moves back in the negative $y$-direction and ends up south of the equator (negative $y$).
The binding energy released in the detonation wave is converted into the kinetic energy 
associated with expansion within $\sim$1 s after the detonation wave completes its passage over 
the star.  The total energy budget is shown in Figure~\ref{fig:energy} for the model ignited
25 km off center.  By $t\sim4$ s the remnant is transitioning into a state of free expansion 
and assumes a self-similar shape which is no longer changing with time and the radial velocity
is well described by a linear dependence on the distance from the center of the remnant.
Axial and equatorial profiles of density and radial velocity in the remnant are presented in 
Figures~\ref{fig:det-profiles} for two models which span the explosion outcomes in our study.

\par Shown in Figure~\ref{fig:dens-contours} is the late time ($t>4$ s) remant shape presented as 
a series of logarithmically spaced density contours for the same two models in Figure~\ref{fig:det-profiles}.  
What can be seen in this figure is that the asymmetry imparted by the off-center ignition and surface detonation
manifests as a shift in the center of density contours even though each individual contour
is well described by a circle. The overall shape of the density distribution, therefore, can 
be characterized by the radius $R_c$ and center $y_c$ of the circles which best describe each 
density contour. This information is shown in Figure~\ref{fig:shape}.  The degree to which these 
two curves ($R_c$ and $y_c$) approximate the remnant is shown by the thin line in 
Figure~\ref{fig:det-profiles}, which is the function $r(\rho) = y_c(\rho)\pm R_c(\rho)$.  

\par Superimposed over the relatively smooth overall shape of the final remnant are smaller scale density 
inhomogeneities due to the turbulent flow associated with the deflagration and the surface flow which 
preceeded detonation. These perturbations are quantified in the bottom panel of Figure~\ref{fig:shape} 
as root mean square (rms) deviations in density taken along the best fit circle at each density.   
The general trend is that more expanded stellar cores have larger
density perturbations in their surface layers at the time
of detonation.  This can be accounted for partly by the fact that more expansion results from
a larger amount of energy liberated in the deflagration which goes into powering the surface flow.  
Additionally, more expanded cores are less stably stratified at their surfaces (lower gravity and 
shallower pressure gradients) and so are more easily perturbed by the surface flow which passes over 
the star before detonation.

\section{NUCLEOSYNTHESIS}
\label{sec:yields}

\par The nucleosynthetic yield for the type of explosion model studied in this paper
consist of a mixture of ash due to both a deflagration and a detonation.  
The total amount of the star burned in the deflagration amounts to less than $\sim$0.1 \msun
with nearly the entire remaining mass of the star consumed by the detonation wave which follows.
The progress variables described in \S\ref{sec:numerics} which are used to parameterize
the compositional evolution and energy release due to nuclear burning allow us to calculate the 
bulk yield of IMEs and NSE material directly from the multi-dimensional
simulation data by performing simple sums \citep{calder2007,townsley2007}. In Table~\ref{tab:models} 
the total mass burned in the deflagration M$_{\rm def}$ and the detonation M$_{\rm det}$ is summarized,
including the budget of IME and NSE material.
The deflagration, propagated with the ADR flame model (\S\ref{sec:numerics}), produces a total yield which 
is approximately one third IMEs and two thirds NSE material, while more than 90\% of the material burned 
in the detonation is completely relaxed to NSE.  

\par A general feature of these explosion models is that higher density cores at the time of detonation 
produce a larger yield of NSE material and a smaller yield of IMEs.  This trend is summarized in 
Figure~\ref{fig:mnse-rhoc} which relates the final NSE yield to the central density of the white dwarf
at the time of detonation.  The total mass of material having a density exceeding $\rho=10^7$ g/cm$^3$
is also shown as a function of central density, and provides a good measure of the mass of material that will 
burn to NSE in the detonation. The relationship between the amount of mass above a certain density and 
the central density is a property of the initial white dwarf density structure 
and the wave form of the pulsation which is excited in the star 
by the deflagration. The dashed line in Figure~\ref{fig:mnse-rhoc} shows the relationship expected if the 
pulsation is described by the fundamental mode of the linear wave equation.  This mode fits the simulation 
data remarkably well considering how large (and non-linear) the pulsation amplitude is for the low central density end of the
figure.  But this can be understood by the fact that the fundamental mode is close to homologous, i.e., the 
displacement is nearly directly proportional to the radius of the white dwarf.

\par The data for the 3D model described in \S\ref{subsec:jet} has been included in 
Figure~\ref{fig:mnse-rhoc} for comparison, and shows that the overall character of the 
expansion driven by the deflagration is not dimensionality dependent, nor is the nucleosynthesis 
that takes place in the detonation.  Data from an additional 3D simulation which burned
significantly more mass in the deflagration is shown and labeled ``3D Multi''. This simulation 
data, which is part of an extended suite of 3D models investigating the mapping of ignition conditions to 
final explosion energies and nucleosynthetic yields (G. C. Jordan IV et al., in prep), was ignited
by a uniform distribution of 30 flame bubbles enclosed in an 80 km radius sphere having its center displaced 100 km
from the stellar center.  
This distribution of ignition points is intended to be more representative
 than single-point ignition of the non-axisymmetric conditions created by
 the interaction of the growing bubble with the pre-ignition convective
 core.
This model demonstrates that fundamental mode radial pulsation is a good description of
the core dynamics preceeding detonation even for significant degrees of expansion.
This model also demonstrates that off-center deflagration models are capable of producing a broad range of 
NSE and \iso{Ni}{56} masses, and not just the most luminous SNe Ia events.

\subsection{Iron Peak Freeze-Out Yields: Method}
\label{subsec:freezeout-method}

\par As NSE material expands and cools in the rarefaction that follows the detonation 
wave, nuclear reactions eventually cease, the composition no longer changes and the 
material is said to have gone through {\em freeze-out}. 
In this section we describe a methodology to efficiently and accurately calculate 
iron peak yields for material which burns to NSE and then freezes out in the expansion following the 
detonation wave.  In our hydrodynamic simulations $\sim10^5$ to $\sim10^6$ Lagrangian tracer 
particles are passively advected through the computational domain with an initial distribution that
evenly samples the underlying mass distribution. Nucleosynthetic yields can then be calcualted by 
integrating nuclear reaction networks along each of these particle trajectories and then summing the 
yields. However, when the large number of particle trajectories required for accurate yield estimates is
multiplied by the number of simulation models desired for study, the computational cost of this
brute force method of post-processing becomes prohibitively expensive. 
Therefore, we have developed an alternative approach to calculate the final composition of
material processed by the detonation, which takes advantage of the fact that the final nucleosynthetic 
yield $X_{i,f}$ of material burned to NSE depends only on the final entropy 
$S_f$, expansion timescale $\tau$, and degree of neutronization $\eta_f$ of the detonated material to a high 
degree of precision with $X_{i,f} = X_{i,f}(S_f,\eta_f,\tau)$.

% I. INDIVIDUAL TRAJECTORY
\subsubsection{Individual Trajectories}

\par The temperature and density of a generic tracer particle processed by the detonation
is presented in Figure~\ref{fig:trajectory}. The evolution of $Y_e$ and the abundances of 
nuclei along this trajectory have been calculated using a nuclear reaction network 
initialized with the initial composition of the white dwarf material in the simulation,
equal mass fractions of \nuc{12}{C} and \nuc{16}{O}. 
The network code used for the  integrations is a version of the network used in 
\citet{calder2007} expanded to 443 nuclear species (see Table~\ref{tab:network}). The 
thermonuclear reaction rates are taken from an expanded version 
~\citetext{Schatz 2005, private communication} of the rate compilation 
REACLIB~\citep{thielemann1986,rauscher2000}.  We have also included the 
temperature-dependent nuclear partition functions provided by~\citet{rauscher2000}, 
both in the determination of the rates of inverse reactions and in our determination 
of NSE abundance patterns. Electron screening of thermonuclear 
reaction rates is incorporated, adopting the relations for weak screening and strong 
screening provided by~\citet{wallace1982} \citep[for additional details see the appendix of][]{calder2007}. 
Contributions from weak reactions are included using the rates provided by \citet{langanke2001}.

\par The time evolution during the rarefaction stage of several abundant
species along this trajectory, parameterized by plasma temperature, is shown in
Figure~\ref{fig:trajectory-burn} (solid lines).  
The NSE composition corresponding to the same density and neutron excess for each temperature along the rarefaction part of the trajectory is also shown for comparison (dashed lines). 
The NSE mass fractions were determined with the NSE solver described in \citet{calder2007} and \cite{seitenzahl2008a}, which uses the same nuclear physics as the reaction network code. 
Figure~\ref{fig:trajectory-burn} illustrates the degree to which adopting an NSE state at a particular ``freeze out temperature'' 
is a poor approximation to the final freeze-out abundances. This is true because nuclei freeze-out 
over a fairly large range in temperature and there is non-trivial evolution in a nuclide's abundance 
after it falls out of NSE but before it reaches its asymptotic freeze-out value.

\par The thermodynamic trajectories for Lagragian elements (i.e., tracer particles) processed 
by the detonation wave are well characterized by an exponential temperature evolution

\begin{equation}
  \label{eqn:temp}
  T(t) = T_0 \exp(-t/\tau)
\end{equation}

\noindent and a corresponding density evolution found by assuming adiabaticity

\begin{equation}
  \label{eqn:entropy}
  S(t) = S(T,\rho, \bar{A}, \bar{Z}) = S_f = (\hbox{constant})
\end{equation}

\noindent where $S_f$ is the final entropy in the post-detonation state and $\bar{A}$ and $\bar{Z}$ are 
the average atomic weight and charge of the plasma during the burn with $Y_e = \bar{Z}/\bar{A}$. 
The density in equation \ref{eqn:entropy} is found using the same Helmholtz equation of state 
\citep{timmes1999,timmes2000a,fryxell2000} used in the hydrodynamic 
simulations. A parameterized trajectory, is shown in Figure~\ref{fig:trajectory}
for comparison to the particle trajectory.  

\par The abundance evolution for this parameterized trajectory is presented in 
Figure~\ref{fig:trajectory-burn} (dotted line) and shows agreement to a high degree 
of precision with the tracer particle trajectory.
Because the composition is well described by a NSE distribution at temperatures above $T_9 \sim 5.5$
the final yields are not dependent on the peak temperature reached by the particle trajectory
and depend only on the entropy, the total amount of neutronization, and the 
expansion timescale. Final asympotic freeze-out yields are safely adopted when the plasma temperature 
drops below $T\sim 10^9$ K with no evolution in the abundances taking place at lower temperatures.  
Because the electron capture rates are a strong function of density, neutronization occurs in the short lived
high density region formed immediately behind the detonation front, while the material is still in NSE and 
well before freeze-out begins.

%\par The stability of the network integrations require that the abundances are initialized with 
%an appropriate NSE distribution, and this is accomplished using the NSE solver described in 
%\citet{calder2007} and \cite{seitenzahl2008a} which uses the same nuclear physics as the 
%reaction network code.
%\par Electron capture rates are a strongly increasing function of density so that most 
%neutronization in the plasma occurs in a narrow region of high density trailing the detonation 
%front while the material is still in NSE, well before freeze out.  This allows us to use a 
%constant electron fraction for our network integrations chosen as the final value in the 
%freeze out.

% II. ENTROPY-NEUTRONIZATION CORRELATION AND LOOKUP TABLE

\subsubsection{Systematic Properties of the Post-Detonation State and Generating a Lookup Table}
\par A tight correlation between the degree of neutronization $\eta_f$ and the final entropy 
 $S_f$ of the detonated material further simplifies the procedure and allows us to calculate 
the final composition for each grid zone in our simulations using a one parameter
freeze-out abundance lookup table $X_{i,f} = X_{i,f}(\eta_f)$.
The asymptotic values of the degree of neutronization $\eta_f = (1-2 Y_{e,f})$
and the entropy $S_f$ for all of the stellar matter burned into NSE in the detonation wave
is found to lie along a narrow ridge in the $S_f$-$\eta_f$ plane as shown
in Figure \ref{fig:ye-s-table}.
This correlation arises from the monotonic dependence of the entropy deposition
and the neutronization rate $\dot{Y_e}(\rho)$ on the post-shock plasma density 
in the detonation wave (which is itself a monotonic function of the pre-shocked, 
upstream density). (Note, this tight correlation doesn't exist for the material which is
burned in the deflagration so that the method described here is not applied to that phase of burning.)

A lookup table is constructed along the black line shown in 
Figure \ref{fig:ye-s-table} and has been sampled at 50 locations logarithmically spaced in neutron excess $\eta$.
For every value of $\eta$ in the table, the freeze-out abundances are calculated by integrating the nuclear reaction network, 
with weak interactions turned off, over an adiabatic analytic trajectories as described above for the corresponding value of 
$S_f$.  The network is initialized with NSE composition at $T_9 = 6.0$ and the integration constinued until $T_9<1.0$, at which 
time freeze-out has occured.  A summary of the stable iron peak yields for material along this locus of points is presented in 
Figure~\ref{fig:yields-table}.  The freeze-out abundances for a computational zone are then be found by using the table entry with 
the corresponding value of $\eta$. The table lookup is computationally fast, and once the table is created no additional network 
calculations are necessary.

As described above, the final freezeout yield depends on the expansion timescale $\tau$.
The expansion timescale, defined by eq.[\ref{eqn:temp}] and found by fitting an exponential to the temperature histories
of the tracer particles, varies smoothly across the face of the white 
dwarf in the narrow range $0.2 < \tau < 0.6$ s for all of the models simulated 
(the range is narrower for an individual explosion model). If one wanted  
to incorporate this information into the processing of the final yields, the range in expansion timescales would need 
to be reflected in the network calculations. 
In the work presented here, we adopt a central value of $\tau = 0.4$ s and 
we discuss the sensitivity of the final yields to variations in this value in \S\ref{subsec:final-yields} below.

% III. PRESENTATION OF FINAL YIELDS
\subsection{Iron Peak Freeze-Out Yields: Results}
\label{subsec:final-yields}

\par We calculate the final yield of stable iron peak isotopes for all of our explosion models
using the procedure outlined in \S\ref{subsec:freezeout-method} above. In Figure~\ref{fig:yields}
we present yields for isotopes in the mass range A = 45 to 68 (\nuc{45}{Ti} to \nuc{68}{Zn}), 
accounting for the decay of radioactive isotopes.  
The yields ($X_i$) are scaled to the \iso{Fe}{56} abundance ($X_{Fe}$, from the decay chain
\nuc{56}{Ni}$\rightarrow$\nuc{56}{Co}$\rightarrow$\nuc{56}{Fe}) and the corresponding relative 
solar system ratio ($X_{i,\odot}/X_{Fe,\odot}$) based on the abundances of \citet{lodders2003}.
In Table~\ref{tab:scaled-yields} we present the elemental abundances for the iron peak elements 
from Ti to Zn scaled to Fe and solar system ratios. In all cases we highlight results for three 
explosion models which bracket the range of initial conditions and final outcomes found in our 
simulation suite. In addition, we provide a detailed list of the final integrated iron peak yields in units
of solar mass in Table~\ref{tab:yields}, including the abundances of radioactive isotopes 
and their half lives.  This table can be used to determine the absolute yield of a particular
isotope, or to examine the isotopic ratios of specific elements of interest.

\par The iron peak yields are similar to pure deflagration
models such as the one-dimensional model of Nomoto \citep[W7 yields in][]{brachwitz2000} and the
three dimensional model presented in \citet{travaglio2004}. 
The iron peak yield for the pure deflagration models are more neutron rich than our 
models, however, because of the higher densities under which the deflagration burns material
to NSE. The highest density core to detonate in our model suite (with $r_{\rm off}$=100 km)
neutronized the most in the detonation and bears the most similarity to a pure deflagration 
model in terms of integrated yields.
The most neutron-rich isotopes of each element (e.g. \iso{Ti}{50}, \iso{Cr}{54},
\iso{Fe}{58}) have no appreciable contribution from the NSE material created by the detonation.  There
is likely some of these species in the small amount of deflagration material not included in our
post-processing.  Notably, none of our models produce untoward overabundances ($\gtrsim 2$ times solar,
indicated by the dotted lines) of either \iso{Fe}{54} or
\iso{Ni}{58}. Overproduction of these nuclides continues to be a serious shortcoming of deflagration
models, both spherical and multi-dimensional, which process much of the stellar interior to NSE before
expansion can occur.
The spatial distribution of the material
in our detonation models, however, remains layered in space and velocity while the burning products
in the pure deflagration model are strongly mixed due to the turbulent nature under which burning proceeds
in a deflagration \citep[e.g.][]{roepke2007b,gamezo2003}. 

\par Interestingly, the total yield of \iso{Ni}{56} in all of the explosion models presented here is
$\sim$1.1\msun  independent of the degree of expansion which takes place prior to detonation.
This is due to a self regulating process comprised of pre-expansion and neutronization which
counteract each other. While the total yield of material burned to NSE is larger for stars
which detonate at higher central densities, more neutronization takes place which shifts 
the iron peak yield to more neutron rich isotopes and away from \iso{Ni}{56} \citep[see e.g.][]{timmes2003}.  
The highest density core at detonation produces overall more stable iron peak isotopes 
but approximately the same \iso{Ni}{56} yield as the core with the lowest density at detonation.
It can be seen from Figure~\ref{fig:mnse-rhoc}, however, that this trend cannot hold for 
significantly more expanded cores since the total mass of high density material drops off 
precipitously as lower central densities are reached and will therefore result in SNe Ia 
explosions which have smaller \iso{Ni}{56} yields.

\par {\em Dependence on progenitor neutronization. ---}
The progenitor white dwarf model used in our explosion calculations 
is composed of equal parts \nuc{12}{C} and \nuc{16}{O} so that $Y_{e} = 0.5$ everywhere in the 
unburned fuel prior to detonation.
However, the progenitor is expected to develop a neutron excess before flame ignition
both during the CNO and He burning cycles and during a $\sim$1000 yr epoch of hydrostatic 
carbon burning which is sometimes referred to as ``simmering''. 
Recent studies of the ``simmering'' epoch indicate that when carbon burning runs away locally
and a flame is born, $\eta^{\rm sim} \approx 10^{-3}$
\citep{piro2008,chamulak2008}, while stars with an initial metallicity comparable to solar 
will develop a neutron excess of $\eta_{\odot} \approx 1.5\times 10^{-3}$  by the time core He burning 
commences. 

\par The neutronization which takes place during the detonation is restricted to the densest, 
central-most regions of the stellar core because of the strong density dependence of the electron 
capture rates (see Figure~\ref{fig:det-ye}). The resulting distribution of neutronization
is presented in Figure~\ref{fig:ye-s-cdf}. This distribution 
extends to values lower than the minimum  $\eta_{\rm min}^{\rm sim}$ expected in the progenitor 
prior to explosive burning.  We explore the impact that such a neutronization 
floor will have on the yields by enforcing $\eta = \max(\eta_{\rm min},\eta)$ prior to 
calculating the iron peak yields using the lookup table method described in \S\ref{subsec:freezeout-method}.
The results are presented in Figure~\ref{fig:yields} (right panel) which shows yields calculated after 
applying neutronization floors of $\eta_{\rm min} = 0$, $10^{-3}$, and $2\times 10^{-3}$ for the model 
which was least neutronized during the detonation (with $r_{\rm off} = $25 km),
and therefore has the most mass affected by a floor in $\eta$.
The iron peak elements  which are primarily produced at low $\eta$ and are therefore most 
strongly affected by a neutronization floor are V, Cr, Mn, and Zn, although the isotopic 
ratios across the entire iron peak are affected.

\par {\em Sensitivity to scatter in $S_f$ and $\tau$. ---} 
It is possible to construct a higher dimensional lookup table for calculating yields 
which accounts for the scatter about the $\eta$-$S_f$ curve used to generate the table 
(Figure~\ref{fig:ye-s-table}), but the total error associated with neglecting this scatter 
in final entropy $S_f$ is small. In Figure~\ref{fig:yields-texp-entropy} (right) we present the total 
variation in the yields due to shifting the $\eta$-$S_f$ curve in final entropy by $\pm$5\%, which is the range of the scatter. 
Similarly, the freezeout timescale that takes place in the wake of the detonation wave
has some scatter about the fiducial value of $\tau=0.4$ s that has been used for the results
presented above, spanning the range $0.2 <\tau < 0.6$ s. 
The total spread in yields adopting the extreme values for $\tau$ is shown in Figure~\ref{fig:yields-texp-entropy} (left). 
The variation in the yield will be significantly smaller than illustrated by these figures since there 
exists a smooth distribution between the extreme values of $S_f$ and $\tau$ 
with the majority of the mass peaked about the central value.  
%While the effects of scatter in 
%$S_f$ and $\tau$ have a small impact on the final yields, it is important to investigate 
%the range of conditions present in any model before applying the freezeout technique described in 
%\S\ref{subsec:freezeout-method} above.

\subsubsection{The velocity distribution of the yield.}
\par {\em The yield of NSE material. ---}
Inferring the abundance stratification in SNe Ia is possible by studying
high fidelity, multi-epoch spectra; a technique which is proving to be a powerful new tool for
constraining explosions models \citep{stehle2005,roepke2007b,mazzali2008}.  The total
number of objects which have had detailed internal abundance stratifications reconstructed
to date is small (only 2 at the time of writing, including 2002bo and 2004eo). These two low 
luminosity SNe have inferred \iso{Ni}{56} masses in the range M[\iso{Ni}{56}]$\sim$0.43 - 0.52 \msun.  
Despite the very different \iso{Ni}{56} masses between these two observed SNe and the explosion models 
presented in this paper, it is interesting to compare the qualitative and quantitative properties of 
the abundance stratifications in an attempt to understand the nature of and the diversity inherent 
in the explosion mechanism.

\par For the case of SN 2004eo, \citet{stehle2005} find M[\iso{Ni}{56}]$\sim$0.43 \msun. In their
reconstruction, the \iso{Ni}{56} mass fraction drops below 0.5 at $v_{\rm exp}\sim$ 7,000 km/s,
and drops below 0.1 at 12,000 km/s.
For the case of SN 2002bo, \citet{mazzali2008} find M[\iso{Ni}{56}]$\sim$0.52 \msun. The \iso{Ni}{56}
mass fraction drops below 0.5 at $v_{\rm exp}\sim$ 10,000 km/s and drops below 0.1 at 15,000 km/s.
In both of these SNe, a high mass fraction of stable Fe (X$_{Fe}\sim 1$) is inferred at low velocities 
$v_{\rm exp} < $ 3,000 km/s.

\par The distribution of the elemental abundances as a function of the 
radial velocity for our explosion models is presented in Figure~\ref{fig:yields-velocity}.
In this figure, the elemental abundances are calculated by summing over isotopes and 
taking into account radioactive decays with half lives less than 1 day.
Two models are shown which bracket the final outcome of all the explosions modeled in this paper. 
Both models produce $\sim$1.1\msun  of \iso{Ni}{56}.
The model ignited with $r_{\rm off}$=25 km is the most expanded at the time of detonation, 
neutronizes the least amount in the detonation, and has the lowest explosion energy, 
E$_{\rm tot} = 1.45\times 10^{51}$ erg.  
The contribution of stable Fe is the smallest in this model, having a mass fraction
X$_{Fe}\sim$10$^{-3}$ out to  $v_{\rm exp}\sim$ 2,000 km/s.  The mass fraction of
\iso{Ni}{56} drops below 0.5 at $v_{\rm exp}\sim$ 14,000 km/s, and drops
below 0.1 at $\sim$ 16,000 km/s.  The model ignited with $r_{\rm off}$=100 km is the least 
expanded at the time of detonation,  is neutronized the most by the detonation wave, and has 
the largest explosion energy, E$_{\rm tot}=1.52\times10^{51}$ erg.
Although the ejecta at low velocities is still dominated by \iso{Ni}{56},
stable Fe with a mass fraction exceeding X$_{Fe}\sim$0.1 is present out to $v_{\rm exp}\sim$6,000 km/s.
The mass fraction of \iso{Ni}{56} drops below 0.5 at a velocity of $v_{\rm exp}\sim$ 16,000 km/s and
drops below 0.1 at 18,500 km/s.

\par While the total yield of \iso{Ni}{56} is significantly larger in the explosion models presented
here, the qualitative layered structure of the remnant and the near absence of unburned
carbon and oxygen are in good agreement between the models and the observations.  As discussed 
in \S\ref{sec:yields} and summarized in Figure~\ref{fig:mnse-rhoc}, lower \iso{Ni}{56}
masses can be produced in surface detonation models which release more energy during the
deflagration phase.  However, detonations which take place at lower central densities 
and produce smaller \iso{Ni}{56} masses, undergo signifiantly less neutronization and will 
therefore fail to reproduce the stable Fe core which has been inferred in the two models 
discussed above.  On the other hand, it is possible that the neutron rich region seen
at low velocities in SNe Ia remnants are a vestige of the progenitor conditions at ignition. 
The nature of the progenitor at flame ignition, including the central density and $Y_e$ 
distribution, are uncertain and will remain so until the evolution leading up to ignition
is better understood, including the much debated and poorly understood Urca process 
\citep[see e.g.,][]{lesaffre2005,arnett1996}.

\par {\em The yield of non-NSE material. ---}
The detailed yield for material which has not completely relaxed to NSE prior to freezeout 
and is composed primarily of IMEs such as Si, S, and Ca is not presented here. In addition to 
producing IMEs, material which has begun 
silicon burning but has not yet reached an NSE state will contribute to the iron peak with isotopic
ratios that are very different from that which reaches an NSE state.
While the impact of this burning process is small for the suite of explosion models presented in this 
paper, which produce primarily NSE material, it is essential to accurately calculate the composition
and distribution of this material for lower luminosity (lower \iso{Ni}{56} mass) SNe Ia explosion models
which will have a significantly larger contribution of non-NSE material.  
Additionaly, although the total contribution of IMEs to the mass of the remnant is small in all of the 
explosions presented here this material plays a central role in modeling the observational 
signatures of these explosion models
and is therefore crucial for comparing our calculations to observational data.
Therefore, a procedure for determining incomplete silicon burning yields which is similar to the 
method described in \S\ref{subsec:freezeout-method} is being developed (C. Meakin et al., in preperation).

%\section{DISCUSSION}
%\par (ala Don Lamb: some words connecting the results of the models presented to the
%observational data. in particular, how the inferred morphology and asymmetries detected
%in spectra and light curves, and the distribution of composition throughout the remantn
%compare with the models presented here.)

\section{CONCLUSIONS}

\par We have studied the final outcomes for a range of single point flame ignition models of 
thermonuclear supernovae within the computational framework developed at the FLASH center
(\S\ref{sec:numerics} and \citet{fryxell2000,calder2007,townsley2007,seitenzahl2008a}).
For the first time in this work, we have extended the 3-stage reactive ash model for nuclear burning
described in \citet{townsley2007}
to study the ignition and propagation of the detonation mode of burning.
As a result, our explosion models
are unique in terms of the degree to which non-idealized nuclear physics are employed, including
a non-static, tabularized treatment of the nuclear statistical equilibrium (NSE) state and the
inclusion of contemporary weak reaction rates \citep{seitenzahl2008a}.
In addition, we have demonstrated here, by reaction network post-processing of 
recorded Lagrangian histories, that the 3-stage reactive ash model
provides a suitable reproduction of fluid density and temperature histories to
allow detailed nucleosynthesis, including self-consistent neutronization.
Using these techniques, we have followed the progression of a thermonuclear flame (deflagration)
from a single ignition point which is varied to successively larger distances from the center of a 
carbon oxygen white dwarf, and we have described in detail the resulting surface flows and 
detonation which ensue.

\par Detonations arise within a colliding surface flow for all models which are 
ignited at a radial location which exceeds $\sim$ 20 km in our 2D simulations.
Flames ignited closer to the stellar center release enough nuclear energy to signifianctly 
expand the stellar core to a degree that it stalls the surface 
flow, thus preventing a strong collision region and detonation.  The nuclear binding energy released 
in these stalled surface flow models, however, is not enough to gravitationally unbind the star and 
they remain viable candidates for a pulsational detonation upon recollapse \citep{khoklov1991,arnett1994}.
Models which detonate release $\sim 2\times10^{51}$ erg in nuclear binding energy, 
resulting in a supernova-like explosion with total energy $E_{\rm tot}\sim1.5\times 10^{51}$ erg.  

\par  In all of the models in our parameter study which produce supernova-like explosions, detonation 
initiates within a jet-like flow which forms in the converging surface flow. This is in agreement with
the results presented in \citet{kasen2007}. However, we do not find that the
detonation initiates through a shock to detonation transition (SDT) as suggested by these authors, but
instead find that the detonation occurs through a gradient mechanism. The initiation of the detonation
takes place within the compressed gas which lies ahead of the high velocity jet, and ahead of the
internal shock which forms within the jet (see \S\ref{sec:det} and Figure~\ref{fig:jet-slice}).  
The focusing of the surface flow and the formation of the jet is also present in 3D simulations
(\S\ref{sec:det}, and \citealt{jordan2007,roepke2007a}),
and is therefore not an artifact of the 2D axisymmetric geometry used.

\par Within a few seconds after the detonation wave disrupts the stellar core, homologous expansion is
beginning to be established. By t$\sim$ 4 s from flame ignition more than 90\% of the total energy 
is in the kinetic energy of expansion, and the expansion velocity has acquired a linear dependence on radius. 
The final remnant posesses both global and small scale asymmetries which will influence the observational signature.
When the remnant enters the homologous expansion phase it is characterized by a smooth, layered 
inner core surrounded by a low density, flocculent layer of deflagration ash which was dumped onto
the surface of the star prior to detonation. The smooth inner core of the remnant has a global north-south 
asymmetry due to off-center ignition and surface detonation, which is well characterized by circular isodensity
contours which are progressively off-center at higher densities (see \S~\ref{subsec:remnant} and Figure~\ref{fig:shape}).

\par We have analyzed in detail the nucleosynthesis of material burned to NSE in the detonation.
These results have been generated from the multi-dimensional simulation data using a newly 
developed post-processing method which takes advantage of the uniqueness of the NSE state and 
systematic properties of detonation waves. The
method, presented in \S\ref{sec:yields}, obviates the need for computationally prohibitive network calculations 
along each of the millions of particle trajectories which are necessary for good mass resolution in 3D explosion models.
This work addresses only material which has relaxed to NSE, which forms, by mass, nearly all of the yield from the
2-dimensional explosion models of this study.
Extending the method to include detailed isotopic yields for material incompletely relaxed to NSE (incomplete carbon,
oxygen, and silicon burning) is being developed and will be described in a forthcoming publication (C. Meakin, in prep).
Nucleosynthesis of material processed in a deflagration instead of detonation burning mode can be processed with a
similarly parameterized method, though requiring more parameters, if it reaches NSE.  This leaves only the relative
minority of tracks in partially burned deflagration material to be processed directly (only a few percent of
all the trajectories).

\par Larger offsets of the ignition point lead to less stellar expansion prior
to detonation and therefore the production of more NSE material.  However, we find that
the amount of \iso{Ni}{56} produced stays roughly fixed at $\sim 1.1 M_\odot$ for all
of our 2-dimensional explosion models which extend down to a
central density of $\rho_c\sim4\times 10^{8}$ g/cm$^{3}$ at the time of detonation.
This regulation is due to the enhanced neutronization at the higher densities characteristic
of the less-expanded cases.
Higher density cores at the time of detonation result in more neutronization, and therefore
a larger fractional yield of stable Fe-peak isotopes (e.g., \iso{Fe}{54} and \iso{Fe}{58}).
The isotopic distribution we find in the Fe-peak is very similar to that found for pure deflagration 
models, but is characterized by a lower degree of neutronization.  Less neutronization is a result of the 
lower densities under which the burning proceeds in our surface detonation models compared to pure
deflagrations, due to the pre-detonation expansion.
Between 0.06 and 
0.14\msun\ of intermediate mass elements are produced at high velocities.
Regions in which more than half of the mass is in the form of IMEs lie
above an expansion velocity of 14,000 km/s for all of the 2-dimensional 
detonation models calculated.

% probably should say more about IMEs compared with observation

\par We successfully reproduced the relationship between the central density and mass-density distribution
in the pre-detonation expanded star by superposing on the hydrostatic star the lowest order radial mode
calculated in a linear approximation.
We find that much smaller \iso{Ni}{56} yields are expected in cores which undergo more expansion prior
to detonation (see Figure~\ref{fig:mnse-rhoc}).  This degree of expansion appears to be achievable in 3-dimensional
simulations which relax the constraints on axisymmetry of the ignition conditions necessary for 2-dimensional
simulations.  Thus it is expected that more realistic simulations, which include the pre-ignition convection
field and its effect on the growing flame bubble, will be characterized by such larger expansions.
However, further analysis of such simulations, which will be the subject of future papers, is required.

% -- I think this is going a bit too far in speculation for the conclusions -- it is fine in the main text -Dean
%In these models, almost no stable Fe will be synthesized
%because detonation will occur at such low densities that the iron peak will be completely dominated by
%\iso{Ni}{56}.  However, the progenitor is thought to undergo
%some degree of neutronization during the simmering phase prior to flame ignition which will result in
%a contribution of stable iron.

\par Future work on elucidating the SNe Ia explosion mechanism which is being pursued at the FLASH center
involves the following. (1) We are extending our survey of the mapping between flame ignition conditions and final outcomes
within the computational framework developed at the FLASH center, including multi-point ignition conditions and 3D models. 
(2) A simulation pipeline is being constructed to generate synthetic observational diagnostics 
for the explosion models, including light curves and spectra, which will allow a more direct comparison between 
the systematic properties of the single degenerate Type Ia model and observational data.

%%%%%%%%%%%%%%%%%%%%%%%%%%%%%%% AKNOWLEDGEMENTS %%%%%%%%%%%%%%%%%%%%%%%%%%%%%%%

\begin{acknowledgements}
  We thank Snezhana Abarzhi for bringing to our attention the literature on
  cummulative jets and shaped charges. We also thank Fang Peng for making her nuclear reaction network code available to us for this work.
  This work is supported in part at the University of Chicago by the Depart of Energy under Grant B523820 to the ASC/Alliances
  Center for Astrophysical Thermonuclear Flashes, and the National Science Foundation under Grant PHY 02-16783 for the 
  Frontier Center ``Joint Institute for Nuclear Astrophysics'' (JINA), and at the Argonne National Laboratory by
  the U.S. Department of Energy, Office of Nuclear Physics, under contract DE-AC02-06CH11357. 
\end{acknowledgements}

%%%%%%%%%%%%%%%%%%%%%%%%%%%%%%%%%%%% APPENDIX %%%%%%%%%%%%%%%%%%%%%%%%%%%%%%%%%

\begin{appendix}

\section{FREEZE OUT ABUNDANCES: DETAILED IRON PEAK YIELDS}

\par The iron peak nucleosynthetic yields for three models spanning the range
of ignition conditions simulated are summarized in Table~\ref{tab:yields}.  
The isotopes presented have been selected based on a limiting abundance 
($M_{i;0,f} > 10^{-20}$ \msun). Two columns are shown for each model including the 
initial yield and the final yield after radioactive decays have been taken into 
account.  The half lives and decay modes are presented for unstable isotopes.

\end{appendix}

%%%%%%%%%%%%%%%%%%%%%%%%%% REFERENCES %%%%%%%%%%%%%%%%%%%%%%%%%%%%%%%%%%
%\input{biblio.tex}

%%%%%%%%%%%%%%%%%%%%%%%%%%%% FIGURES %%%%%%%%%%%%%%%%%%%%%%%%%%%%%%%%%%%
\clearpage

\begin{figure}
  \includegraphics{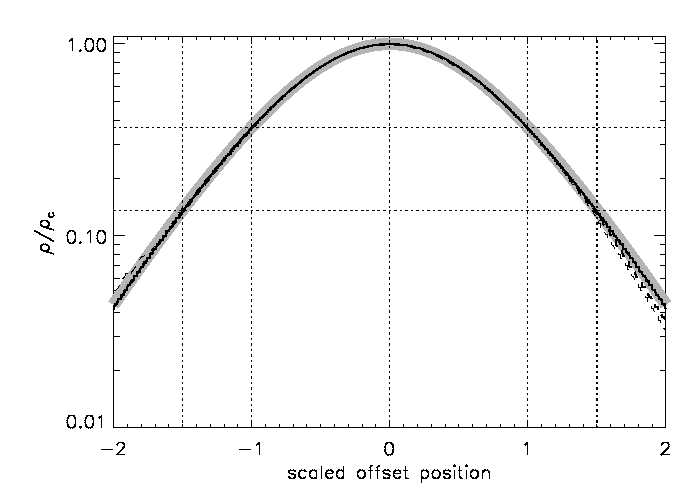}
  \caption{The radial density profile scaled to the central density and the
  density e-folding height for the initial white dwarf (thick grey
  line), and at the time when a surface detonation initiates ($t_{\rm det}$,
  see Table \ref{tab:models}) for flame bubbles ignited  at 25 (thin black lines)
  and 100 km (thick black lines) off-center.
  Equatorial (solid lines) and polar (dashed lines) profiles are shown
  for both of the pre-detonation models and are well described by homologous
  expansion with minimal asymmetry.
    \label{fig:homologous-deflagration}}
\end{figure}

\begin{figure}
  \includegraphics[scale=0.6]{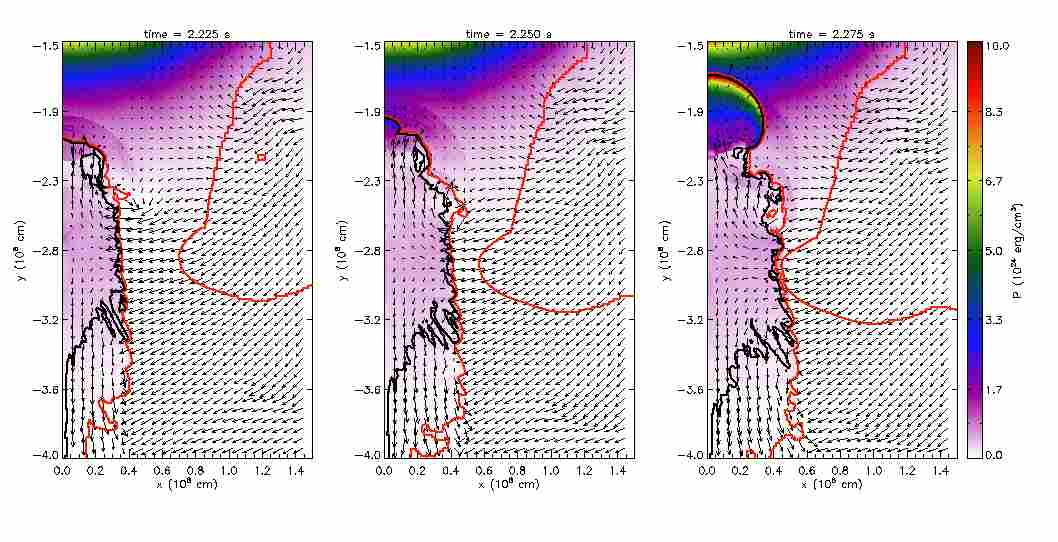}
  \caption{A time sequence showing the ``collision region'' and the bidirectional
    jet-like flow for the model ignited 40 km off-center.
    The pressure field and the velocity field are shown as the detonation wave initiates 
    and begins to break away from the end of the inwardly moving jet component.
    The red and black contour lines indicate carbon depletion at the 1\% and 99\% levels, 
    respectively. In the region where the detonation initiates the
    density and temperature are 10$^7$ g/cm$^3$ and 4$\times$10$^9$ K. The longest
    velocity vector indicates a flow speed of $v_{\rm vec} = 10^9$ cm/s, while other
    vectors have lengths linearly proportional to the flow speed.
    \label{fig:jet}}
\end{figure}

\begin{figure}
  \epsscale{1.0}
  \includegraphics[scale=0.62]{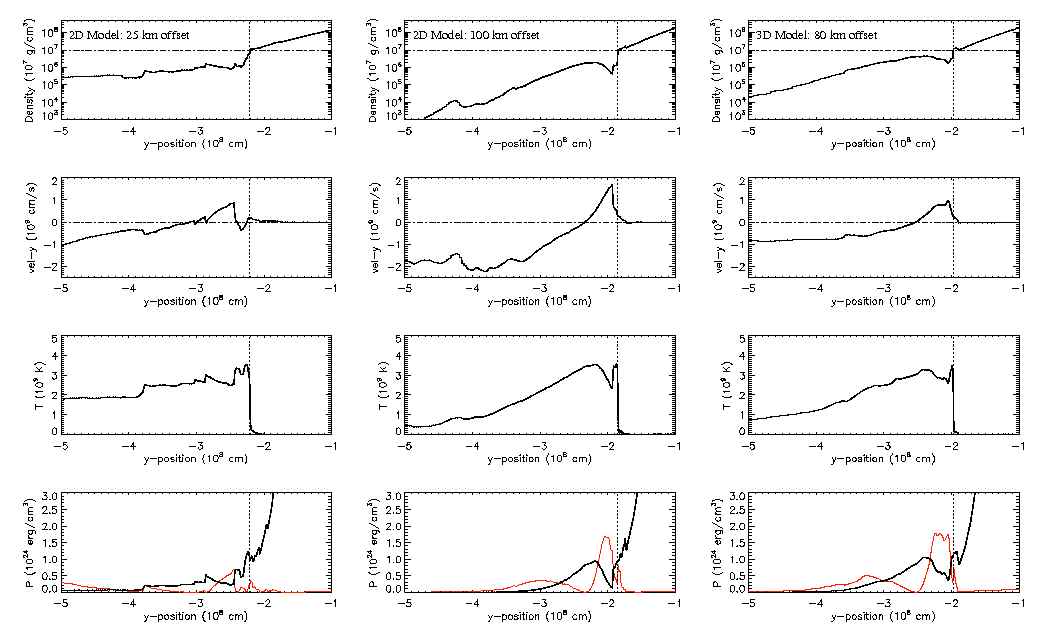}
  \caption{Flow properties along the jet axis prior to detonation for 2D and 3D models, including
    density, radial velocity, temperature, and gas pressure.  The radially directed ram pressure, 
    $p_{\rm ram} = \rho v_r^2$, is shown in the pressure figure for each model by the thin red line.
    The 2D models shown were ignited by 16 km radius flame bubbles with offset (left) $r_{\rm off} = 25$ km and 
    (middle) $r_{\rm off} = 100$ km.  A low resolution ($\Delta = 8$ km) 3D model is shown (right) 
    which was ignited by a 16 km radius flame bubble with offset $r_{\rm off} = 80$ km for comparison.
    The dashed vertical lines mark the locations of the burning front for each model, taken to be
    where $\phi_1 = 0.5$. In the top panel, the horizontal dot-dashed line marks a density of 
    $10^7$ g/cm$^3$, a value above which detonation readily arises in the simulations once it reaches
    a temperature of $T\sim 2\times 10^9$ K. The velocity zero point is marked by a dot-dashed horizontal line.
    \label{fig:jet-slice}}
\end{figure}

\begin{figure}
  \includegraphics[scale=0.6]{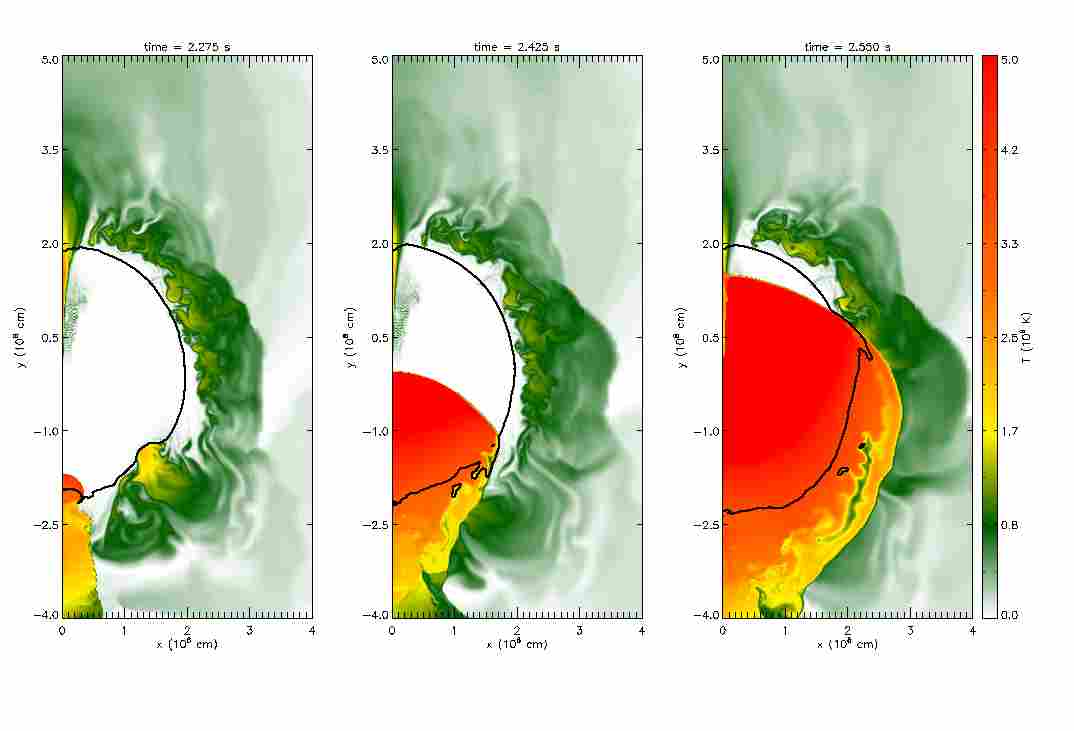}
  \caption{In this time sequence, the detonation wave breaks away from the jet in which it 
    formed and sweeps across the stellar core.  The black line is the 10$^7$ g/cm$^3$ iso-density 
    contour which is roughly coincident with the stellar surface.  
    \label{fig:det-temp}}
\end{figure}

\begin{figure}
  \includegraphics[scale=0.6]{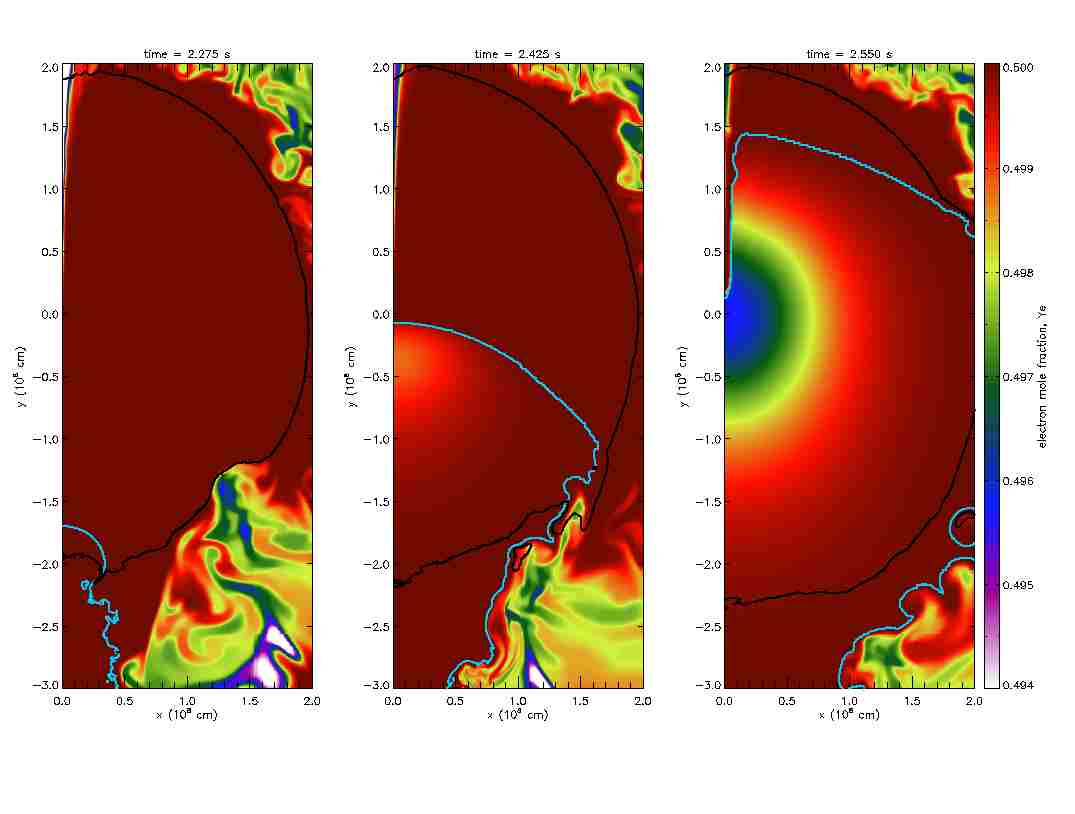}
  \caption{The time evolution of the electron mole fraction ($Y_e$) is shown as the detonation 
    wave passes over the stellar center.  The dip in $Y_e$ reveals that neutronization is taking 
    place in the high density core where material has burned to NSE.  The expansion 
    which follows the detonation freezes the $Y_e$ distribution as the material evolves into a 
    supernova remnant. The neutron rich material (low $Y_e$) which surrounds the stellar
    core is the ash from the deflagration which had burned at high densities before it
    erupted from the star and spread out over the surface.
    The black line is the 10$^7$ g/cm$^3$ iso-density contour and the light 
    blue line indicates the contour of carbon depletion at the 99\% level.\label{fig:det-ye}}
\end{figure}

\begin{figure}
  \includegraphics[scale=0.45]{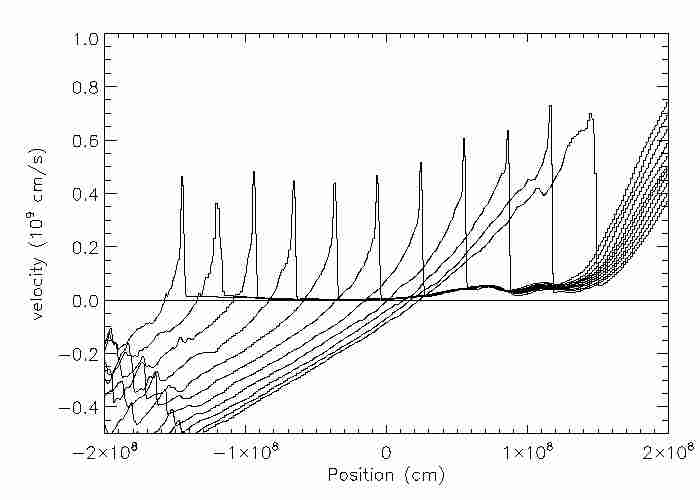}
  \includegraphics[scale=0.45]{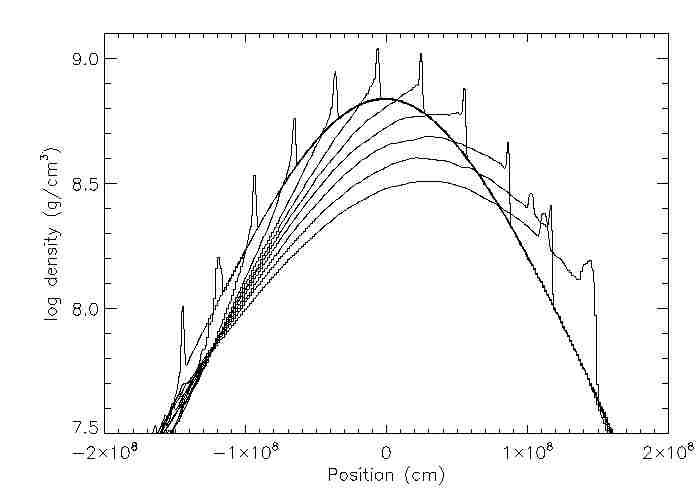}
  \caption{The profile of the (left) velocity and (right) density along the
    polar axis is shown for several moments evenly spaced in time ($\delta t=$0.025 s)  
    as the detonation wave passes across the stellar core for the model ignited
    with a flame bubble 40 km from the stellar center.
    \label{fig:det-propagation}}
\end{figure}

\begin{figure}
  \epsscale{1.0}
  \includegraphics{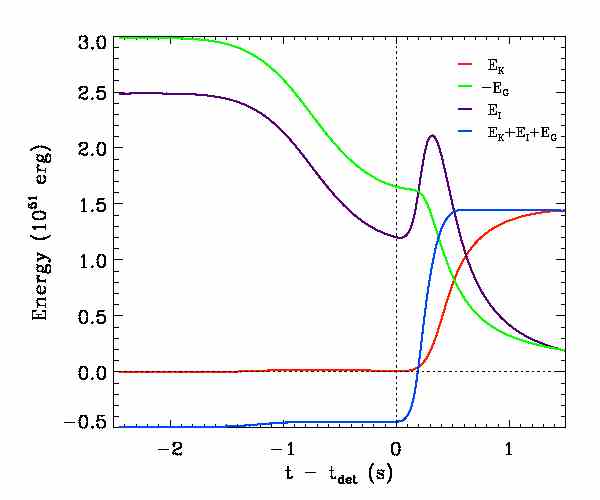}
  \epsscale{1.0}
  \caption{The time evolution of the kinetic, internal, and  gravitational potential 
    energy for a model with a flame bubble ignited 25 km from the stellar center.
    The nuclear energy released by burning in the deflagration ($t < t_{det}$, with $t_{det} = 2.45$s) 
    and the detonation ($t > t_{det}$) can be seen as a change in the sum of 
    these three energy components (blue).\label{fig:energy}}
\end{figure}

\begin{figure}
  \includegraphics[scale=0.7]{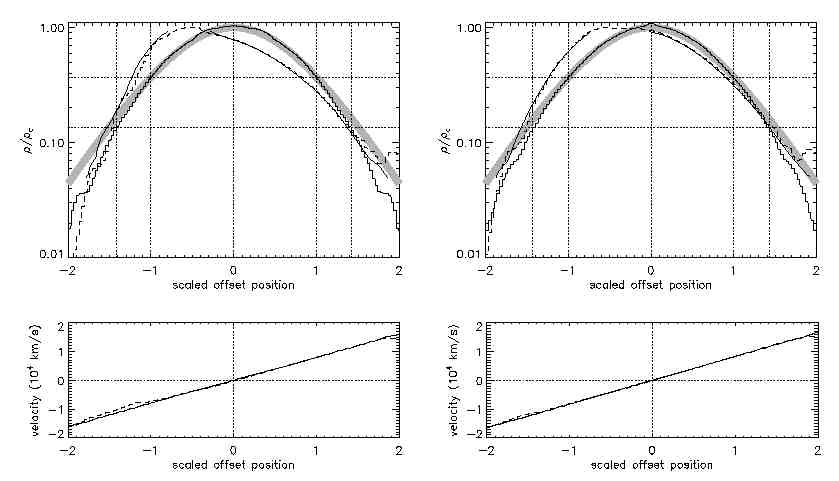}
  \caption{Late time ($t>4$ s) density and velocity profiles for post detonation state models
    having ignition:
    (left) 25 km and (right) 100 km off-center. The density
    is scaled by the peak value and the position is scaled by the density e-folding
    distance in the equatorial direction.
%    The spatial scales for the 25 km and 100 km offset models
%    are  $r_{\rho} = 1.41\times 10^9$ cm and  $r_{\rho} = 1.62\times 10^9$ cm, 
%    respectively.
    The thick gray line shows the scaled density profile 
    of the initial white dwarf model, while the post detonation state model is shown 
    by thick black lines for profiles along the: (solid) equatorial, and (dashed) polar axes.
    The thin black line shows the reconstructed density profile along the polar
    axis based on the contour fits presented in Figure~\ref{fig:shape} (see text for more details).
    \label{fig:det-profiles}}
\end{figure}

%\begin{figure}
%  \epsscale{1.0}
%  \includegraphics{figs/line-density-25and100.ps}
%  \caption{Scaled line density, $\lambda/\lambda_{\rm max}$ (see eq.[\ref{eq:line-density}]), along
%    the symmetry axis at the following times: (black) just prior to detonation, and 
%    (gray) at late times ($t>4$ s).
%    Shown are the models with flame bubbles ignited: (thick) 25 km, and 
%    (thin) 100 km off-center.
%    The spatial scales are the same density e-folding lengths as in Figure~\ref{fig:det-profiles}.
%    \label{fig:line-density}}
%\end{figure}

\begin{figure}
  \epsscale{1.0}
  \includegraphics[scale=0.7]{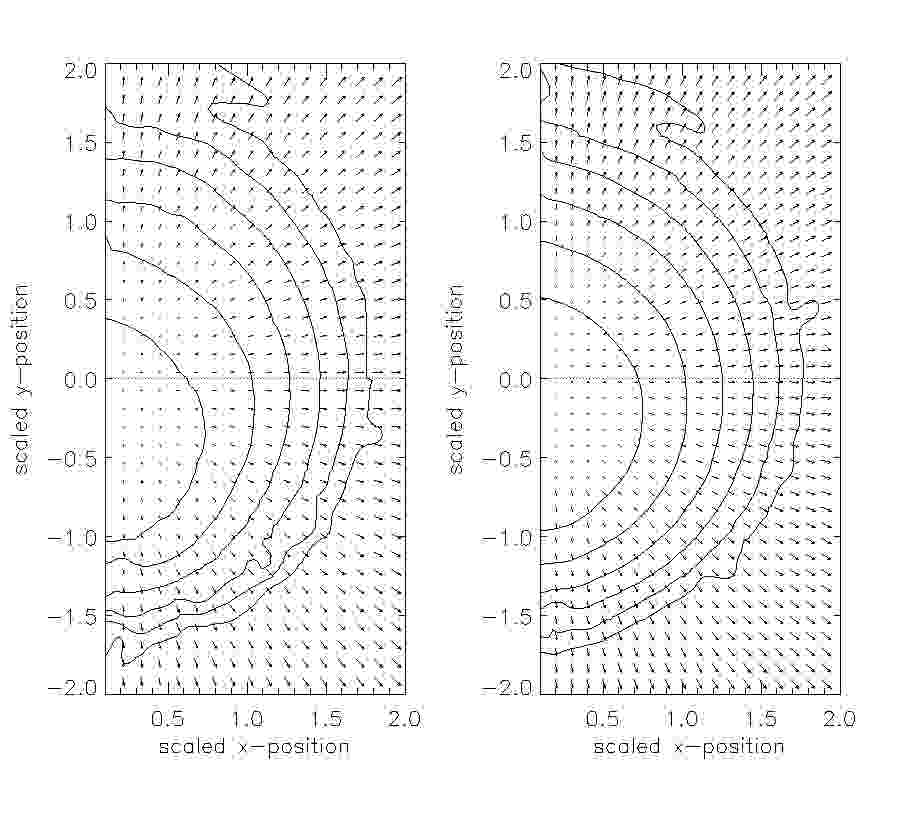}
  \caption{Late time ($t>4$ s) density contours for the models ignited: (left) 25 km, and
    (right) 100 km off-center.  The contours mark the locations at which
    $\ln (\rho/\rho_c) = -0.5, -1, -1.5, -2, -2.5$ where $\rho_c$ is the peak density.
    The magnitude of the largest velocity vectors are (left) $v = 2.4\times 10^9$ cm/s, and 
    (right) $v = 2.5\times 10^9$ cm/s.
    The spatial scales are the same density e-folding lengths as in Figure~\ref{fig:det-profiles}.
    \label{fig:dens-contours}}
\end{figure}

\begin{figure}
  \epsscale{0.8}
  \includegraphics{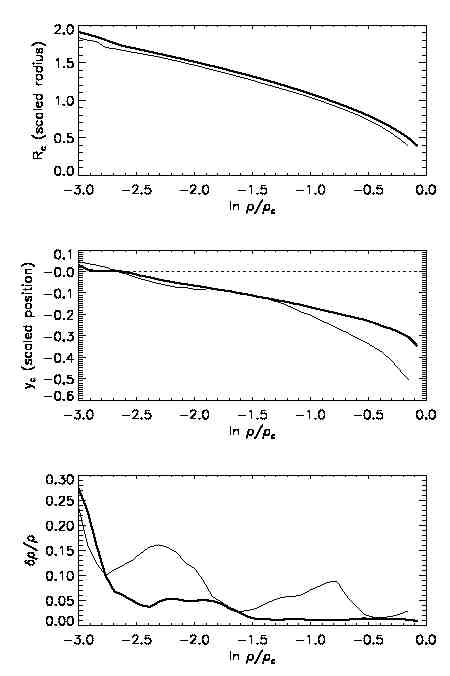}
  \caption{The iso-density contours of the remnant during late times 
    ($t>4$ s, Figure~\ref{fig:dens-contours}) are well described by 
    circles of radius $R_c$ which have centers that are offset from the origin
    by an amount $y_c$ along the symmetry axis. The best fit radii and offsets are
    shown here as a function of density contour value scaled to the central density 
    for the models ignited 25 km (thin-line), and 100 km (thick-line) off-center
    in the panels above: (top) radius, $R_c$; (middle) circle center, $y_c$. 
    The spatial dimensions are scaled in terms of the e-folding density scale
    height in the equatorial direction of the remnant.
    (bottom) The degree of clumpiness is characterized by the ratio of the r.m.s. 
    deviation in density along the best fit circle to the density contour value,
    denoted $\delta\rho/\rho$.
    The density perturbations at high density ($\ln\rho/\rho_c$>1.5) are due primarily
    to the narrow trail of ash left behind as the flame bubble rises our of the
    stellar core.
    \label{fig:shape}}
\end{figure}

\clearpage

\begin{figure}
  \epsscale{1.0}
  \includegraphics{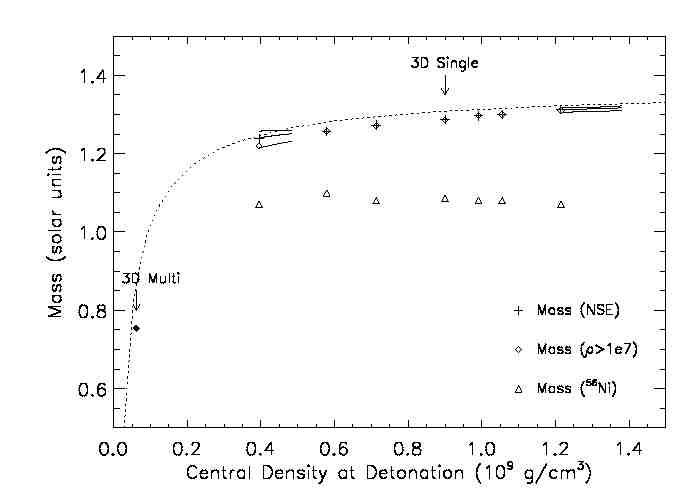}
  \caption{The total mass of NSE material and \nuc{56}{Ni} created in the detonation and the 
    total mass of high density ($\rho>10^7$ g/cm$^3$) matter at detonation is plotted against 
    the central density at detonation for all of the 2D models studied. 
    The total mass of material having a density which exceeds 
    $\rho=5\times 10^6$, $7.5\times 10^6$, and $10^7$ g/cm$^3$ during a 0.25 s time period 
    preceeding detonation is shown by the curves which terminate at the detonation
    density for the 25 km and 100 km off-center ignition models.
    Data points for two 3D models are shown for comparison: the data points labeled 
    ``3D Single'' show the \nuc{56}{Ni}, NSE, and high density material masses for a 
    3D model ignited  80 km  off-center.  The data point labeled ``3D Multi'' shows 
    the central density and the high density material mass for a 3D multi-point ignition 
    model which is described in the text (see \S\ref{sec:yields}).  The dashed line shows 
    the relationship between central density and high density material for the initial
    white dwarf model expanded by the (linear) fundamental pulsation mode.
    \label{fig:mnse-rhoc}}
\end{figure}

\begin{figure}
  \includegraphics[scale=0.6]{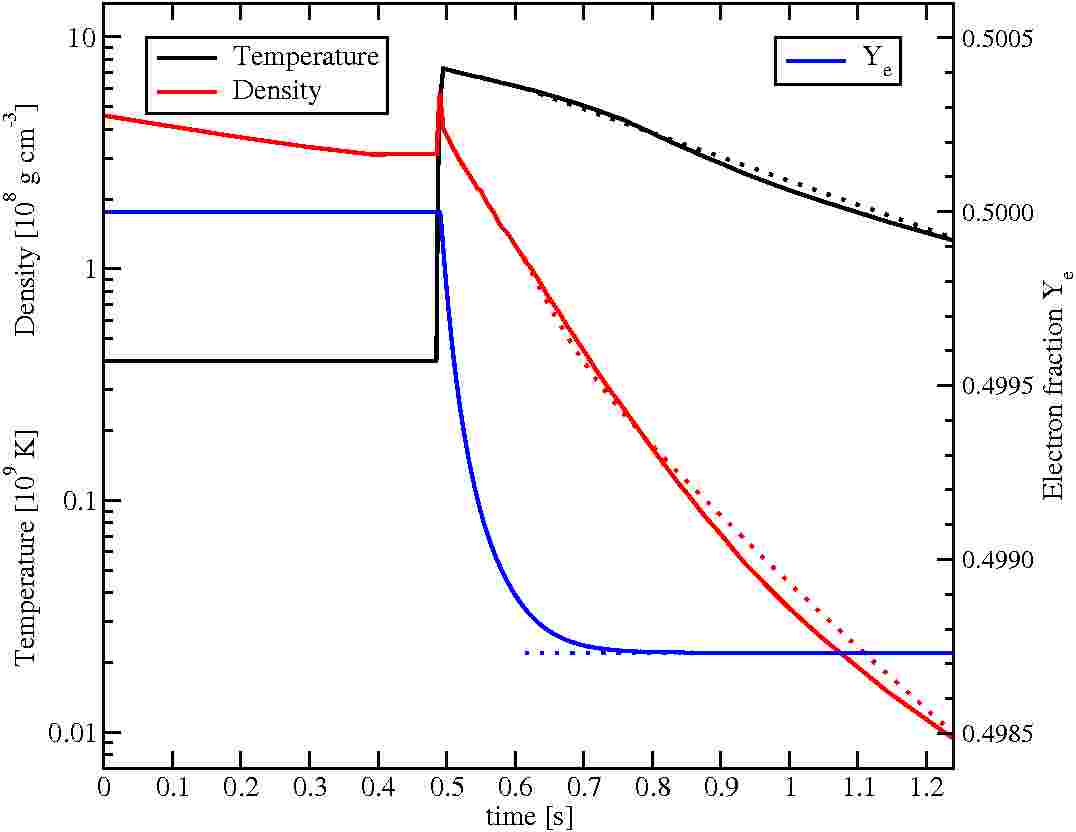}
  \caption{The thermodynamic trajectory of a Lagrangian tracer particle having
    an expansion timescale $\tau = 0.42$s, and final entropy $s$ = 2.273 N$_A k$ and final
    electron mole fraction $Y_e$ = 0.49873.  The dotted line shows the analytic 
    adiabatic fit  to this trajectory, parameterized by $\tau$, $Y_e$, and $s$.
    \label{fig:trajectory}}
\end{figure}

\begin{figure}
  \epsscale{0.9}
  \includegraphics[scale=0.6]{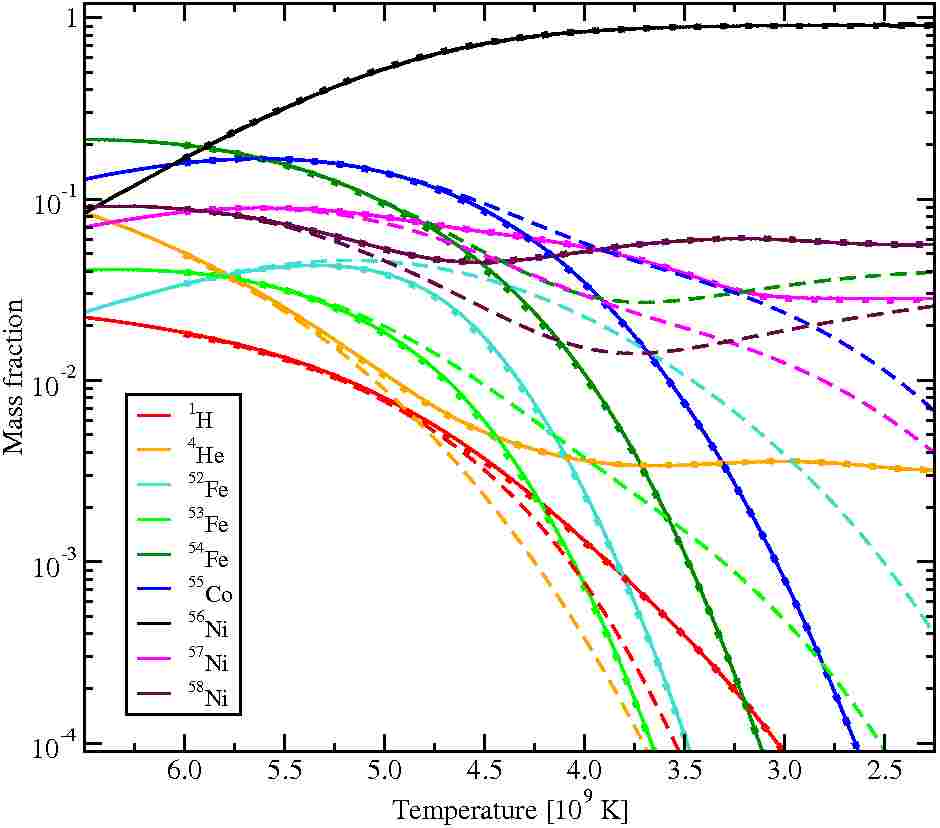}
  \caption{The time evolution is shown for various high abundance iron peak isotopes 
    during the expansion which is here parameterized by the plasma temperature.
    For each species the abundances have been calculated with a network using the thermodynamic 
    trajectory of the tracer particle (solid) and the analytic fit (dotted). For comparison, the
    NSE values are shown (dashed) using the thermodynamic conditions at each point along 
    the particle trajectory.  All three are in good agreement until $T\sim5.5\times10^9$ K, 
    below which the NSE distribution begins to deviate from the network calculation at various 
    temperatures. The tracer particle and the analytic fit agree to a high level of precision 
    through freeze out. \label{fig:trajectory-burn}}
\end{figure}

\begin{figure}
  \includegraphics[scale=0.6]{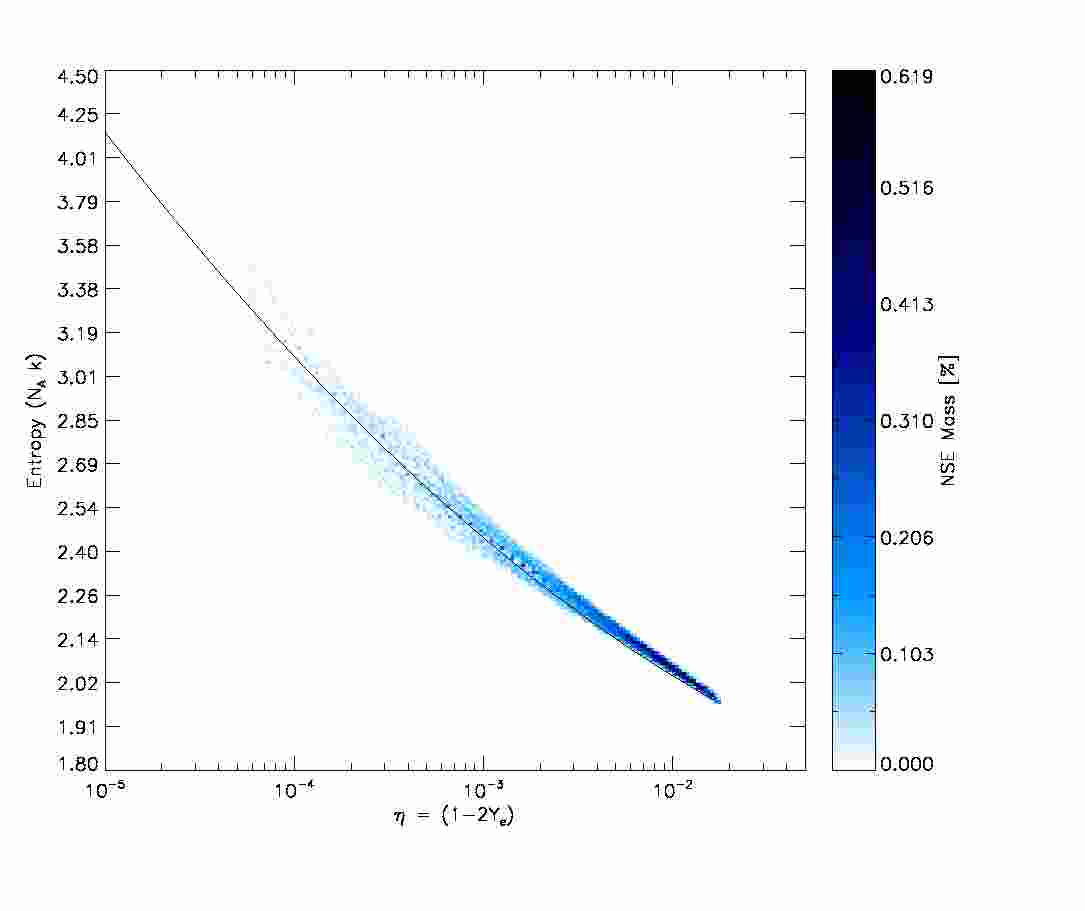}
  \caption{The distribution of NSE mass in entropy and degree of neutronization 
    for the  model with 100 km off-center bubble ignition.  The black line indicates the 
    curve on which the freeze out yield table was calculated (\S\ref{subsec:freezeout-method}).
    %(right) Cumulative distribution function showing the fraction of NSE material below 
    %a certain degree of neutonization.
    \label{fig:ye-s-table}}
\end{figure}

\begin{figure}
  \includegraphics{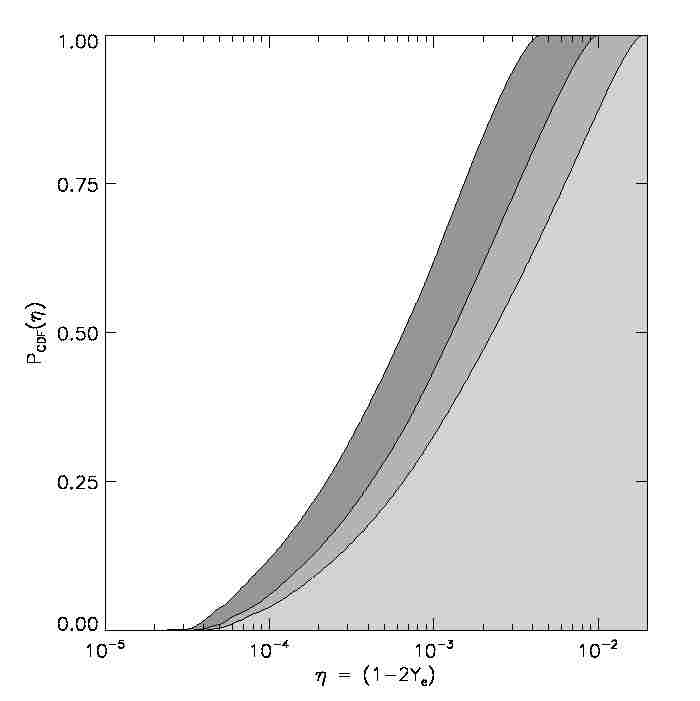}
  \caption{Cumulative distribution functions showing the fraction of NSE material below 
    a certain degree of neutonization for the models ignited: 25 km, 40km, and 100 km 
    off-center (from darkest to lightest, respectively).
    \label{fig:ye-s-cdf}}
\end{figure}

\begin{figure}
  \includegraphics[scale=0.45]{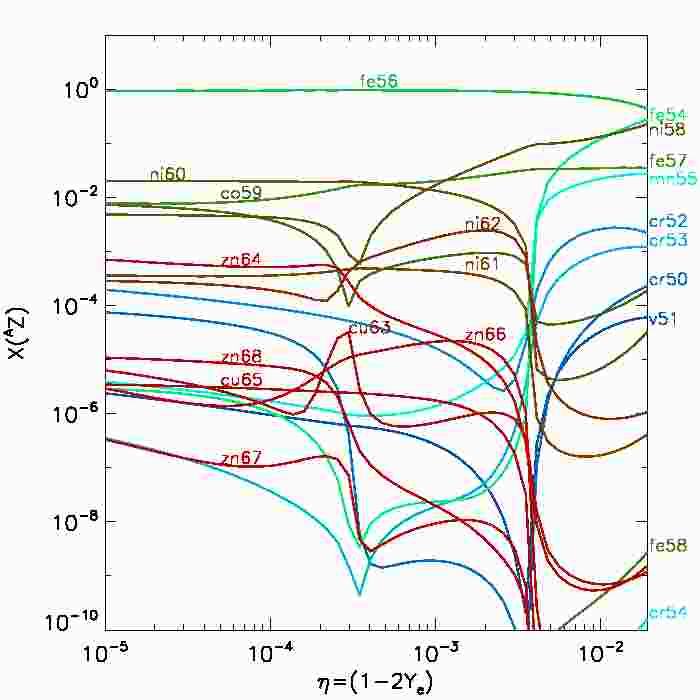}
  \includegraphics[scale=0.45]{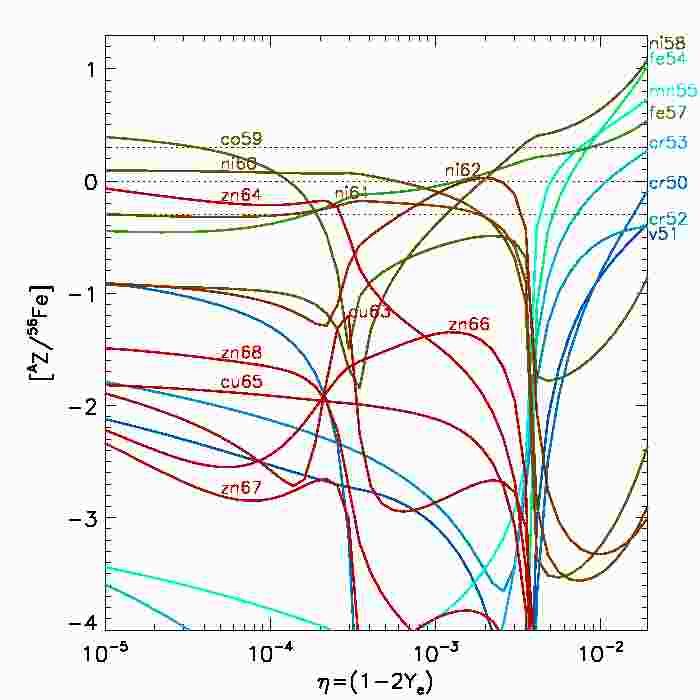}
  \caption{Iron peak freezeout yields, accounting for 
    radioactive decay, as a function of the degree of neutronization.
    The corresponding entropy of the material is related to the degree of neutronization, 
    $S_f = S_f(\eta_f)$, by the black line shown in Figure~\ref{fig:ye-s-table}
    and an expansion timescale of $\tau = 0.4$ s.
    (left) Isotope mass fractions, X(\nuc{A}{Z}) for nuclide with atomic number A and 
    proton number Z.
    (right) Isotope mass fractions scaled to X(\nuc{56}{Fe}) and normalized to the 
    corresponding solar system abundance ratios of \citet{lodders2003} where
    [\nuc{A}{Z}/\nuc{56}{Fe}] =$\log_{10}$ 
    (X(\nuc{A}{Z})/X(\nuc{56}{Fe})) - $\log_{10}$(X(\nuc{A}{Z})/X(\nuc{56}{Fe}))$_{\odot}$.
    The dashed horizontal lines indicate where the scaled abundance ratio is equal to 0.5, 
    1.0, and 2.0 times the solar system value.
    \label{fig:yields-table}}
\end{figure}

\begin{figure}
  \includegraphics[scale=0.45]{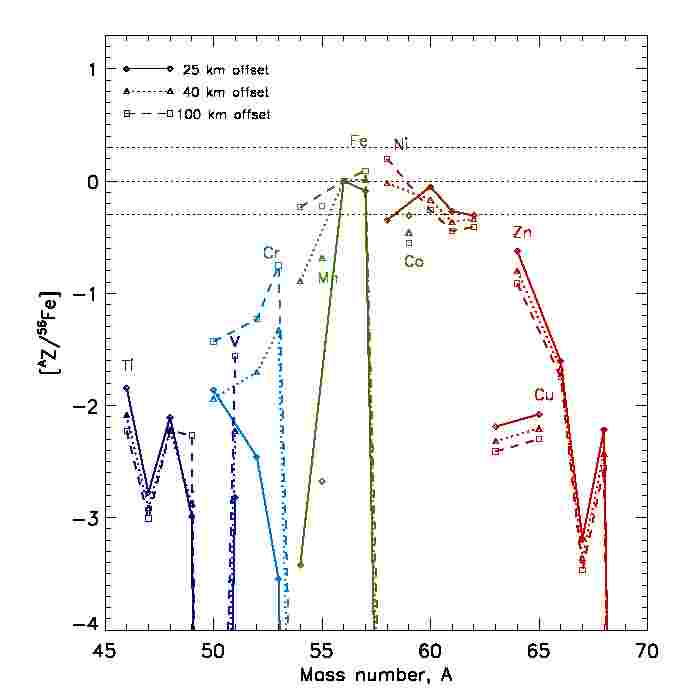}
  \includegraphics[scale=0.45]{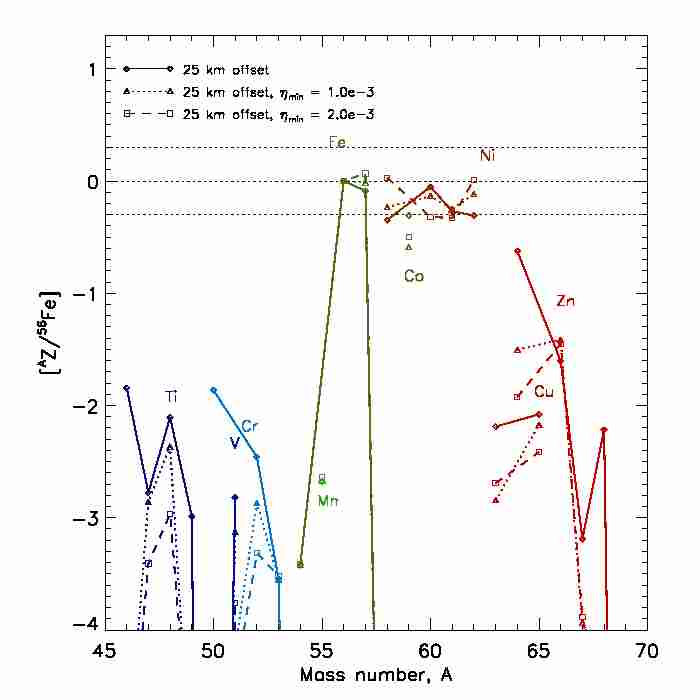}
  \caption{(left) The nucleosynthetic yields normalized by the solar system abundances are 
    shown for three models which span the degree of pre-expansion and neutronization 
    in our simulation suite.
    (right) The variation in the yields due to imposing a neutronization floor of 
    $\eta_{\rm min}$ = 0, $10^{-3}$, and $2\times 10^{-3}$ for the model ignited 
    25 km off-center. Same notation as Figure~\ref{fig:yields-table} \label{fig:yields}}
\end{figure}

\begin{figure}
  \includegraphics[scale=0.45]{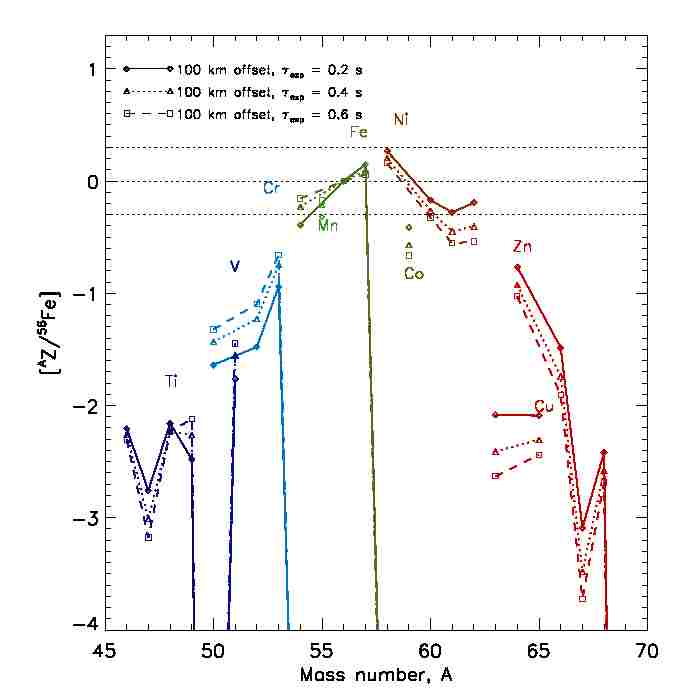}
  \includegraphics[scale=0.45]{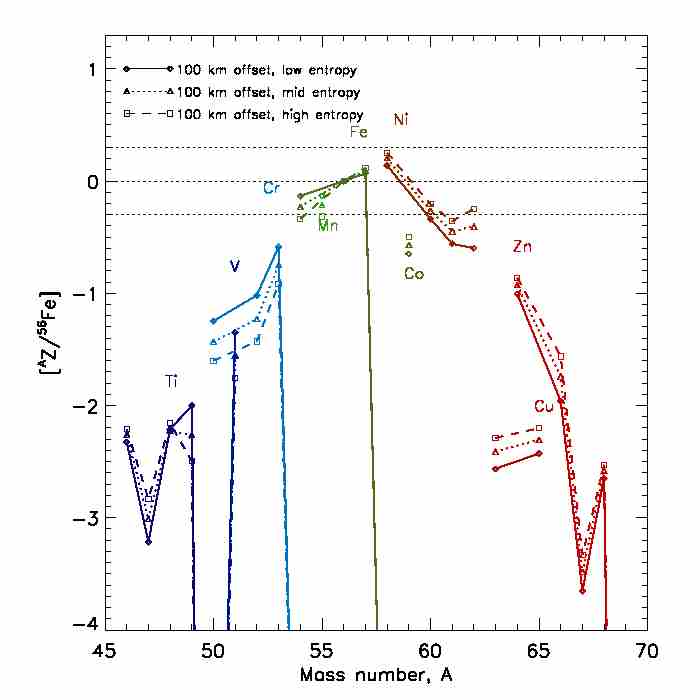}
  \caption{The variation in nucleosynthetic yields due to:
    (left) variations in expansion timescale, with $\tau_{exp} = $0.2, 0.4, and 0.6 s; 
    and (right) entropy changes, where $\pm$5\% variations about the fiducial 
    entropy-neutronization curve, $S_f = S_f(\eta_f)$, shown in Figure~\ref{fig:ye-s-table} 
    have been used. Same notation
    as Figure~\ref{fig:yields-table}.
    \label{fig:yields-texp-entropy}}
\end{figure}

\begin{figure}
  \includegraphics[scale=0.45]{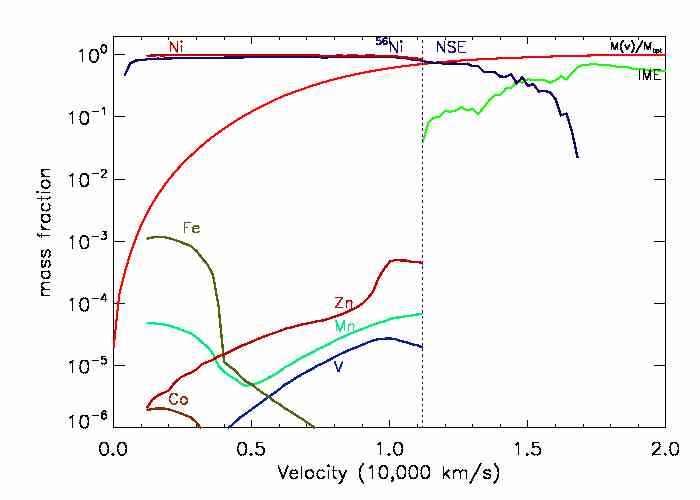}
  \includegraphics[scale=0.45]{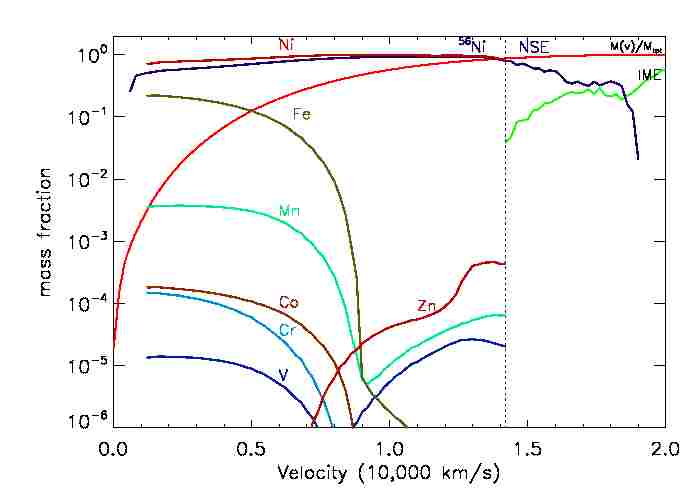}
  \caption{Distribution of elemental abundances are shown as a function of 
    expansion velocity for models having initial flame bubbles ignited (left) 
    25 km and (right) 100 km from the stellar center. The elemental yields are 
    calculated by taking into account radioactive decays with half lives less than 
    1 day.  The dotted vertical line indicates the velocity above which less than 
    95\% of the material has been burned to NSE and therefore our nucleosynthesis 
    post-processing method (\S\ref{subsec:freezeout-method}) is no longer reliable. 
    At velocities where less than 95\% of the material burns to NSE we show only 
    the total fraction of NSE and intermediate mass elements (IMEs).  The red 
    curve shows the fraction of the total stellar mass interior to the 
    velocity. \label{fig:yields-velocity}}
\end{figure}

%%%%%%%%%%%%%%%%%%%%%%%%% TABLES %%%%%%%%%%%%%%%%%%%%%%%%%%%%%%%%%%
\clearpage

\input{tab1.tex}

\input{tab2.tex}

\input{tab3.tex}

\input{tab4.tex}

\end{document}

%% file: tab1.tex
%%%%%%%%%%%%%%%%%%%%%%%%%%%%%%%%%%%%%
% MODEL SUMMARY TABLE
%%%%%%%%%%%%%%%%%%%%%%%%%%%%%%%%%%%%%

\begin{deluxetable}{cccccccccccc}
%\tabletypesize{\scriptsize}
\tablecaption{Simulation Model Parameters\label{tab:models}}
\tablehead{
  \mcol{1}{c}{$r_{\rm off}$} &
  \mcol{1}{c}{$t_{\rm det}$} &
  \mcol{1}{c}{$r_{\rm det}$} &
  \mcol{1}{c}{E$_{\rm n,def}$} &
  \mcol{1}{c}{M$_{\rm def}$} &
  \mcol{1}{c}{M$_{\rm def}^{\rm IME}$} &
  \mcol{1}{c}{M$_{\rm def}^{\rm NSE}$} &
  \mcol{1}{c}{M$_{\rm det}^{\rm IME}$} &
  \mcol{1}{c}{M$_{\rm det}^{\rm NSE}$} &
  \mcol{1}{c}{M[$^{56}$Ni]\tnm{a}} &
  \mcol{1}{c}{E$_{\rm tot}$\tnm{c}}\\
  \mcol{1}{c}{(km)} &
  \mcol{1}{c}{(s)} &
  \mcol{1}{c}{(10$^3$ km)} &
  \mcol{1}{c}{(10$^{49}$ erg)} &
  \mcol{1}{c}{(\msun)} &
  \mcol{1}{c}{(\msun)} &
  \mcol{1}{c}{(\msun)}&
  \mcol{1}{c}{(\msun)} &
  \mcol{1}{c}{(\msun)} &
  \mcol{1}{c}{(\msun)} &
  \mcol{1}{c}{10$^{51}$ erg}}
\startdata
20  & \dots  & \dots        & \dots   & \dots    &  \dots     &  \dots     & \dots & \dots & \dots & \dots  \\
25  & 2.45   & 2.19         & 5.73    & 3.88(-2) & 1.50(-2)   & 2.38(-2)   & 0.127 & 1.239 & 1.07  & 1.445 \\
30  & 2.32   & 2.05\tnm{b}  & 4.91    & 2.96(-2) & 1.01(-2)   & 1.96(-2)   & 0.110 & 1.257 & 1.10  & 1.447 \\
40  & 2.24   & 2.04         & 3.84    & 2.76(-2) & 1.20(-2)   & 1.56(-2)   & 0.090 & 1.273 & 1.08  & 1.469 \\
60  & 1.99   & 1.84         & 2.86    & 1.84(-2) & 0.59(-2)   & 1.24(-2)   & 0.071 & 1.296 & 1.08  & 1.498 \\
80  & 1.98   & 1.83         & 2.64    & 1.70(-2) & 0.62(-2)   & 1.08(-2)   & 0.068 & 1.299 & 1.08  & 1.500 \\
100 & 1.89   & 1.79         & 2.22    & 1.38(-2) & 0.49(-2)   & 0.89(-2)   & 0.058 & 1.312 & 1.07  & 1.517 \\
\enddata
\tnt{a}{The quoted \iso{Ni}{56} mass is that created in the detonation.}
\tnt{b}{off-axis detonation: ($x,y$)=(-2.04, 0.191)$\times$10$^8$ cm.}
\tnt{c}{The total energy at t=3.0 s.}
\end{deluxetable}

%% file: tab2.tex
%%%%%%%%%%%%%%%%%%%%%%%%%%%%%%%%%%%%%%%%%%%%%%%%%%%%%%%%%%
%  TABLE: NUCLEAR REACTION NETWORK USED FOR SIMULATION   %
%%%%%%%%%%%%%%%%%%%%%%%%%%%%%%%%%%%%%%%%%%%%%%%%%%%%%%%%%%

\begin{deluxetable}{lccc}
  \tabletypesize{\scriptsize} \tablewidth{0pt} 
  \tablecaption{443 Nuclei Included in Network Calculations\label{tab:network}} 
  \tablehead{
    Element &
    \mcol{1}{c}{Z} &
    \mcol{1}{c}{A$_{\rm min}$}&
    \mcol{1}{c}{A$_{\rm max}$}}

  \startdata
n   &  0   &    1   &    1  \\
H   &  1   &    1   &    1  \\
He  &  2   &    3   &    4  \\
C   &  6   &    12  &    12 \\
O   &  8   &    16  &    22 \\
F   &  9   &    16  &    24 \\
Ne  &  10  &    16  &    26 \\
Na  &  11  &    18  &    30 \\
Mg  &  12  &    19  &    30 \\
Al  &  13  &    22  &    33 \\
Si  &  14  &    23  &    36 \\
P   &  15  &    26  &    38 \\
S   &  16  &    27  &    40 \\
Cl  &  17  &    30  &    42 \\
Ar  &  18  &    31  &    45 \\
K   &  19  &    34  &    50 \\
Ca  &  20  &    35  &    52 \\
Sc  &  21  &    39  &    57 \\
Ti  &  22  &    39  &    59 \\
V   &  23  &    43  &    61 \\
Cr  &  24  &    44  &    61 \\
Mn  &  25  &    48  &    62 \\
Fe  &  26  &    48  &    68 \\
Co  &  27  &    51  &    66 \\
Ni  &  28  &    52  &    70 \\
Cu  &  29  &    55  &    72 \\
Zn  &  30  &    56  &    74 \\
Ga  &  31  &    60  &    74 \\
Ge  &  32  &    62  &    76 \\
As  &  33  &    67  &    78 \\
Se  &  34  &    67  &    82 \\
Br  &  35  &    71  &    81 \\
Kr  &  36  &    71  &    86 \\
\enddata

\end{deluxetable}

%% file: tab3.tex
%%%%%%%%%%%%%%%%%%%%%%%%%%%%%%%%%%%%%%%%%%%%%%%%%%%%%%%%%%%%%%%%%
%  TABLE: SOME FE-PEAK YIELDS SCALED TO SOLAR SYSTEM ABUNDANCES
%%%%%%%%%%%%%%%%%%%%%%%%%%%%%%%%%%%%%%%%%%%%%%%%%%%%%%%%%%%%%%%%%

\begin{deluxetable}{cccc}
  \tabletypesize{\scriptsize} \tablewidth{0pt} 
  \tablecaption{Scaled elemental Fe-peak yields, \solrat{X}{Fe}, for 
    select explosion models.\tablenotemark{a}
    \label{tab:scaled-yields}}
  \tablehead{
    \mcol{1}{c}{$^{\rm Z}$X} &
    \mcol{1}{c}{$r_{\rm off}=$25 km} &
    \mcol{1}{c}{$r_{\rm off}=$40 km} &
    \mcol{1}{c}{$r_{\rm off}=$100 km}}
\startdata
\mcol{1}{r}{\iso{Ti}{46,47,48,49,50}}  & -2.12  & -2.27 & -2.29 \\
\mcol{1}{r}{\iso{V}{50,51}}            & -2.79  & -2.21 & -1.57 \\
\mcol{1}{r}{\iso{Cr}{50,52,53,54}}     & -2.43  & -1.65 & -1.18 \\
\mcol{1}{r}{\iso{Mn}{55}}              & -2.65  & -0.67 & -0.23 \\
\mcol{1}{r}{\iso{Co}{59}}              & -0.28  & -0.44 & -0.56 \\
\mcol{1}{r}{\iso{Ni}{58,60,61,62,64}}  & -0.22  & -0.06 &  0.08 \\
\mcol{1}{r}{\iso{Cu}{63,65}}           & -2.12  & -2.26 & -2.38 \\
\mcol{1}{r}{\iso{Zn}{64,66,67,68,70}}  & -0.89  & -1.07 & -1.20 \\

%\mcol{1}{r}{\iso{Ti}{46,47,48,49,50}}  & -2.119  & -2.266 & -2.287 \\
%\mcol{1}{r}{\iso{V}{50,51}}            & -2.793  & -2.210 & -1.565 \\
%\mcol{1}{r}{\iso{Cr}{50,52,53,54}}     & -2.426  & -1.648 & -1.177 \\
%\mcol{1}{r}{\iso{Mn}{55}}              & -2.646  & -0.670 & -0.227 \\
%\mcol{1}{r}{\iso{Co}{59}}              & -0.279  & -0.442 & -0.561 \\
%\mcol{1}{r}{\iso{Ni}{58,60,61,62,64}}  & -0.220  & -0.055 &  0.083 \\
%\mcol{1}{r}{\iso{Cu}{63,65}}           & -2.122  & -2.263 & -2.379 \\
%\mcol{1}{r}{\iso{Zn}{64,66,67,68,70}}  & -0.888  & -1.069 & -1.203 \\

%\mcol{1}{r}{\iso{Ti}{46,47,48,49,50}}  & \mcol{3}{c}{\dots/-2.119/\dots}   & \mcol{3}{c}{\dots/-2.287/\dots} \\
%\mcol{1}{r}{\iso{V}{50,51}}            & \mcol{3}{c}{\dots/-2.793/\dots}   & \mcol{3}{c}{\dots/-1.565/\dots} \\
%\mcol{1}{r}{\iso{Cr}{50,52,53,54}}     & \mcol{3}{c}{\dots/-2.426/\dots}   & \mcol{3}{c}{\dots/-1.177/\dots} \\
%\mcol{1}{r}{\iso{Mn}{55}}              & \mcol{3}{c}{\dots/-2.646/\dots}   & \mcol{3}{c}{\dots/-0.227/\dots} \\
%\mcol{1}{r}{\iso{Co}{59}}              & \mcol{3}{c}{\dots/-0.279/\dots}   & \mcol{3}{c}{\dots/-0.561/\dots} \\
%\mcol{1}{r}{\iso{Ni}{58,60,61,62,64}}  & \mcol{3}{c}{\dots/-0.220/\dots}   & \mcol{3}{c}{\dots/ 0.083/\dots} \\
%\mcol{1}{r}{\iso{Cu}{63,65}}           & \mcol{3}{c}{\dots/-2.122/\dots}   & \mcol{3}{c}{\dots/-2.379/\dots} \\
%\mcol{1}{r}{\iso{Zn}{64,66,67,68,70}}  & \mcol{3}{c}{\dots/-0.888/\dots}   & \mcol{3}{c}{\dots/-1.203/\dots} \\
\enddata

\tablenotetext{a}{Here the total elemental abundance is the sum over the listed 
  stable isotopes \iso{X}{\rm Z} and is denoted X = $\sum$($^{\rm z}$X). The standard notation for the 
  logarithmic abundance ratio relative to the solar system abundance ratio is used where
  \solrat{X}{Fe} = $\log_{10}$(X/Fe) - $\log_{10}$(X/Fe)$_{\odot}$.}
\end{deluxetable}

% FIDUCAL DATA
% ------------
%
% 25 km
%
% Ti =   -2.119
%  V =   -2.793
% Cr =   -2.426
% Mn =   -2.646
% Co =   -0.279
% Ni =   -0.220
% Cu =   -2.122
% Zn =   -0.888
%
% 40 km
%
% Ti =   -2.266
%  V =   -2.210
% Cr =   -1.648
% Mn =   -0.670
% Co =   -0.442
% Ni =   -0.055
% Cu =   -2.263
% Zn =   -1.069
%
% 100 km
%
% Ti =   -2.287
%  V =   -1.565
% Cr =   -1.177
% Mn =   -0.227
% Co =   -0.561
% Ni =    0.083
% Cu =   -2.379
% Zn =   -1.203

%% file: tab4.tex
%%%%%%%%%%%%%%%%%%%%%%%%%%%%%%%%%%%%%%%
%  TABLE: Yields for 2D GCD Models    %
%%%%%%%%%%%%%%%%%%%%%%%%%%%%%%%%%%%%%%%

\begin{deluxetable}{lcccccccc}
  \tabletypesize{\scriptsize} 
  \tablewidth{0pt} 
  \tablecaption{Integrated Iron Peak Yields (\msun)\label{tab:yields}}
  \tablehead{
    \mcol{1}{c}{Nuclide} &
    \mcol{1}{c}{half life}&
    \mcol{1}{c}{decay mode} &
    \mcol{2}{c}{$r_{\rm off}$=25 km} &
    \mcol{2}{c}{$r_{\rm off}$=40 km} &
    \mcol{2}{c}{$r_{\rm off}$=100 km}\\
    \mcol{1}{c}{}&
    \mcol{1}{c}{$\tau_{1/2}$}&
    \mcol{1}{c}{\betm/\betp}&
    \mcol{1}{c}{$M_i^0$}&
    \mcol{1}{c}{$M_i^f$} &
    \mcol{1}{c}{$M_i^0$}&
    \mcol{1}{c}{$M_i^f$} &
    \mcol{1}{c}{$M_i^0$}&
    \mcol{1}{c}{$M_i^f$}}

\startdata 
% \iso{Cr}{46} & 0.26 s   & \betp &  &   &  2.53e-11  & \dots \\
  %\iso{Mn}{49} & 382 ms   & \betp &  &   &  4.06e-12  & \dots \\
  %\iso{Ni}{54} & 104 ms   & \betp &  &   &  1.00e-13  &  \dots \\
  \iso{Ti}{43} & 509 ms   & \betp & 9.78e-08 & \dots	    &7.45e-08 & \dots	   &5.93e-08 & \dots	      \\
  \iso{Ti}{44} & 60.0 y   & \betp & 1.01e-05 & \dots	    &7.22e-06 & \dots	   &5.66e-06 & \dots	      \\
  \iso{Ti}{45} & 184.8 m  & \betp & 3.09e-10 & \dots	    &3.17e-10 & \dots	   &6.84e-10 & \dots	      \\
  \iso{Ti}{46} & \dots    & \dots & 2.70e-11 & 3.27e-06     &4.29e-10 & 1.89e-06   &6.03e-09 & 1.36e-06    \\
  \iso{Ti}{47} & \dots    & \dots & 3.92e-15 & 3.50e-07     &1.02e-14 & 2.60e-07   &1.30e-13 & 2.07e-07    \\
  \iso{Ti}{48} & \dots    & \dots & 6.64e-16 & 1.67e-05     &7.65e-16 & 1.29e-05   &5.07e-15 & 1.28e-05    \\
  \iso{Ti}{49} & \dots    & \dots &  \dots    & 1.64e-07    & \dots   & 2.16e-07   &1.37e-20 & 8.66e-07    \\
  \iso{V}{46}  & 422.5 ms & \betp & 4.34e-10 & \dots	    &3.86e-10 & \dots	   &3.23e-10 & \dots	      \\
  \iso{V}{47}  & 32.6 m   & \betp & 2.66e-08 & \dots	    &2.29e-08 & \dots	   &2.09e-08 & \dots	      \\
  \iso{V}{48}  & 15.97 d  & \betp & 6.12e-13 & \dots	    &2.11e-12 & \dots	   &1.80e-11 & \dots	      \\
  \iso{V}{49}  & 329 d    & \betp & 2.30e-14 & \dots	    &2.72e-13 & \dots	   &3.64e-12 & \dots	      \\
  \iso{V}{51}  & \dots    & \dots & 3.85e-20 & 5.49e-07     &1.91e-18 & 2.19e-06   &4.61e-17 & 1.01e-05    \\
  \iso{Cr}{47} & 500 ms   & \betp & 3.24e-07 & \dots	    &2.37e-07 & \dots	   &1.86e-07 & \dots	      \\
  \iso{Cr}{48} & 21.56 h  & \betp & 1.67e-05 & \dots	    &1.29e-05 & \dots	   &1.28e-05 & \dots	      \\
  \iso{Cr}{49} & 42.3 m   & \betp & 3.29e-10 & \dots	    &1.23e-07 & \dots	   &7.99e-07 & \dots	      \\
  \iso{Cr}{50} & \dots    & \dots & 6.85e-10 & 9.47e-06     &2.61e-06 & 8.10e-06   &2.19e-05 & 2.58e-05    \\
  \iso{Cr}{51} & 27.7 d   & \betp & 7.25e-14 & \dots	    &3.61e-10 & \dots	   &3.82e-09 & \dots	      \\
  \iso{Cr}{52} & \dots    & \dots & 8.35e-15 & 4.83e-05     &7.62e-11 & 2.80e-04   &1.11e-09 & 8.19e-04    \\
  \iso{Cr}{53} & \dots    & \dots &  \dots    & 4.53e-07    &3.29e-17 & 7.63e-05   &7.14e-16 & 2.85e-04    \\
  \iso{Cr}{54} & \dots    & \dots &  \dots    & 2.06e-16    & \dots   & 6.06e-13   &1.29e-19 & 7.17e-12    \\
  \iso{Mn}{50} & 283.88 ms& \betp & 8.38e-10 & \dots	    &9.03e-10 & \dots	   &1.31e-09 & \dots	      \\
  \iso{Mn}{51} & 46.2 m   & \betp & 1.38e-08 & \dots	    &1.82e-06 & \dots	   &9.84e-06 & \dots	      \\
  \iso{Mn}{52} & 5.591 d  & \betp & 2.40e-11 & \dots	    &2.63e-08 & \dots	   &1.66e-07 & \dots	      \\
  \iso{Mn}{53} & 3.74 My  & \betp & 1.12e-11 & \dots	    &2.24e-08 & \dots	   &1.87e-07 & \dots	      \\
  \iso{Mn}{54} & 312.12 d & \betp & 2.06e-16 & \dots	    &6.06e-13 & \dots	   &7.17e-12 & \dots	      \\
  \iso{Mn}{55} & \dots    & \dots & 7.21e-19 & 2.63e-05     &2.41e-15 & 2.60e-03   &3.72e-14 & 7.54e-03    \\
  \iso{Fe}{50} & 155 ms   & \betp & 9.47e-06 & \dots	    &5.48e-06 & \dots	   &3.94e-06 & \dots	      \\
  \iso{Fe}{51} & 305 ms   & \betp & 5.36e-07 & \dots	    &3.71e-07 & \dots	   &2.86e-07 & \dots	      \\
  \iso{Fe}{52} & 8.275 h  & \betp & 4.83e-05 & \dots	    &2.80e-04 & \dots	   &8.19e-04 & \dots	      \\
  \iso{Fe}{53} & 8.51 m   & \betp & 4.38e-07 & \dots	    &7.63e-05 & \dots	   &2.85e-04 & \dots	      \\
  \iso{Fe}{54} & \dots    & \dots & 2.45e-05 & 2.49e-05     &8.50e-03 & 8.50e-03   &3.85e-02 & 3.85e-02    \\
  \iso{Fe}{55} & 2.737 y  & \betp & 5.79e-09 & \dots	    &2.63e-06 & \dots	   &1.56e-05 & \dots	      \\
  \iso{Fe}{56} & \dots    & \dots & 1.21e-10 & 1.07e+00     &6.82e-08 & 1.08e+00   &5.38e-07 & 1.07e+00    \\
  \iso{Fe}{57} & \dots    & \dots & 2.61e-17 & 2.05e-02     &1.97e-14 & 2.62e-02   &2.45e-13 & 3.11e-02    \\
  \iso{Fe}{58} & \dots    & \dots & 1.67e-20 & 2.48e-13     &1.09e-17 & 2.57e-11   &1.40e-16 & 1.60e-10    \\
%  \iso{Fe}{59} & 44.503 d & \betm & \dots    & \dots        &\dots    & \dots      &\dots    & \dots       \\
%  \iso{Fe}{60} & 1.5 My   & \betm & \dots    & \dots        &\dots    & \dots      &\dots    & \dots       \\
  \iso{Co}{54} & 192.23 ms& \betp & 4.02e-08 & \dots	    &9.85e-08 & \dots	   &1.56e-07 & \dots	      \\
  \iso{Co}{55} & 17.53 h  & \betp & 2.54e-05 & \dots	    &2.59e-03 & \dots	   &7.52e-03 & \dots	      \\
  \iso{Co}{56} & 77.23 d  & \betp & 1.20e-07 & \dots	    &1.13e-05 & \dots	   &4.08e-05 & \dots	      \\
  \iso{Co}{57} & 271.74 d & \betp & 1.20e-08 & \dots	    &1.42e-06 & \dots	   &6.34e-06 & \dots	      \\
  \iso{Co}{58} & 70.86 d  & \betp & 2.48e-13 & \dots	    &2.57e-11 & \dots	   &1.60e-10 & \dots	      \\
  \iso{Co}{59} & \dots    & \dots & 1.54e-15 & 1.67e-03     &1.90e-13 & 1.20e-03   &1.20e-12 & 9.55e-04    \\
%  \iso{Co}{60} & 5.2714 y & \betm & \dots    & \dots        &\dots    & \dots      &3.49e-20 & \dots	      \\
  \iso{Ni}{55} & 202 ms   & \betp & 9.36e-07 & \dots	    &6.82e-07 & \dots	   &5.38e-07 & \dots	      \\
  \iso{Ni}{56} & 6.075 d  & \betp & 1.07e+00 & \dots	    &1.08e+00 & \dots	   &1.07e+00 & \dots	      \\
  \iso{Ni}{57} & 35.6 h   & \betp & 2.05e-02 & \dots	    &2.62e-02 & \dots	   &3.11e-02 & \dots	      \\
  \iso{Ni}{58} & \dots    & \dots & 1.96e-02 & 2.10e-02     &4.45e-02 & 4.54e-02   &7.38e-02 & 7.44e-02    \\
  \iso{Ni}{59} & 76 ky    & \betp & 5.26e-07 & \dots	    &1.11e-05 & \dots	   &2.91e-05 & \dots	      \\
  \iso{Ni}{60} & \dots    & \dots & 3.26e-08 & 1.66e-02     &1.01e-06 & 1.28e-02   &2.97e-06 & 1.03e-02    \\
  \iso{Ni}{61} & \dots    & \dots & 6.04e-14 & 4.42e-04     &1.90e-12 & 3.62e-04   &6.26e-12 & 2.95e-04    \\
  \iso{Ni}{62} & \dots    & \dots & 8.78e-17 & 1.31e-03     &2.90e-15 & 1.23e-03   &1.16e-14 & 1.05e-03    \\
%  \iso{Ni}{63} & 100.1 y  & \betm & \dots    & \dots        &\dots    & \dots      &\dots    & \dots       \\
  \iso{Cu}{58} & 7.0 s    & \betp & 1.38e-03 & \dots	    &8.78e-04 & \dots	   &6.55e-04 & \dots	      \\
  \iso{Cu}{59} & 0.46 s   & \betp & 5.30e-04 & \dots	    &5.14e-04 & \dots	   &4.40e-04 & \dots	      \\
  \iso{Cu}{60} & 0.57 s   & \betp & 8.17e-07 & \dots	    &9.25e-07 & \dots	   &8.49e-07 & \dots	      \\
  \iso{Cu}{61} & 0.27 s   & \betp & 6.44e-08 & \dots	    &1.32e-07 & \dots	   &1.66e-07 & \dots	      \\
  \iso{Cu}{62} & 0.19 s   & \betp & 7.53e-11 & \dots	    &7.42e-11 & \dots	   &6.89e-11 & \dots	      \\
  \iso{Cu}{63} & \dots    & \dots & 4.39e-14 & 3.69e-06     &1.43e-13 & 2.77e-06   &2.28e-13 & 2.23e-06    \\
  \iso{Cu}{65} & \dots    & \dots &  \dots    & 2.19e-06    & \dots   & 1.66e-06   & \dots    & 1.32e-06   \\
  \iso{Zn}{59} & 182 ms   & \betp & 1.14e-03 & \dots	    &6.73e-04 & \dots	   &4.86e-04 & \dots	      \\
  \iso{Zn}{60} & 2.38 m   & \betp & 1.66e-02 & \dots	    &1.28e-02 & \dots	   &1.03e-02 & \dots	      \\
  \iso{Zn}{61} & 89.1 s   & \betp & 4.42e-04 & \dots	    &3.62e-04 & \dots	   &2.95e-04 & \dots	      \\
  \iso{Zn}{62} & 9.186 h  & \betp & 1.25e-03 & \dots	    &1.20e-03 & \dots	   &1.02e-03 & \dots	      \\
  \iso{Zn}{63} & 38.47 m  & \betp & 2.86e-07 & \dots	    &2.97e-07 & \dots	   &2.60e-07 & \dots	      \\
  \iso{Zn}{64} & \dots    & \dots & 3.28e-09 & 2.25e-04     &4.33e-09 & 1.52e-04   &4.10e-09 & 1.16e-04    \\
  \iso{Zn}{66} & \dots    & \dots & 6.17e-18 & 1.39e-05     &5.82e-17 & 1.22e-05   &7.54e-17 & 1.01e-05    \\
  \iso{Zn}{67} & \dots    & \dots &  \dots    & 5.41e-08    & \dots   & 3.70e-08   & \dots    & 2.85e-08   \\
  \iso{Zn}{68} & \dots    & \dots &  \dots    & 2.35e-06    & \dots   & 1.46e-06   & \dots    & 1.08e-06   \\
\enddata
%\tablecomments{--}
\end{deluxetable}